\documentclass[10pt, twocolumn]{article}
 
\title{\textbf{AI Adoption in NGOs: A Systematic Literature Review}}

\author{
    \fontsize{11}{13}\selectfont 
    1\ts{st} Janne Rotter \\
    \fontsize{10}{11}\selectfont 
    Universitat Pompeu Fabra\\
    \fontsize{10}{11}\selectfont 
    Barcelona, Spain\\
    \fontsize{10}{11}\selectfont   jannedavid.rotter01@estudiant.upf.edu\\
    \fontsize{10}{11}\selectfont 
    ORCID: 
    \href{https://orcid.org/0009-0001-3520-958X}{ 0009-0001-3520-958X}
    \and
    \fontsize{11}{13}\selectfont 
    2\ts{nd} William Bailkoski \\
    \fontsize{10}{11}\selectfont 
    Universitat Pompeu Fabra\\
    \fontsize{10}{11}\selectfont 
    Barcelona, Spain\\
    \fontsize{10}{11}\selectfont   william.bailkoski01@estudiant.upf.edu\\
    \fontsize{10}{11}\selectfont 
    ORCID: 
    \href{https://orcid.org/0009-0009-0744-6228}{ 0009-0009-0744-6228}
}

\date{}

\usepackage{lipsum}
\usepackage{lmodern} 
\usepackage{array} 
\usepackage{booktabs} 
\usepackage{dblfloatfix}

\usepackage[
layout=letterpaper, 
paper=letterpaper, 
portrait, 
head=0.5in,
foot=0.5in,
top=1in, 
bottom=1in, 
left=0.75in, 
right=0.75in
]{geometry}
\usepackage{tabularx}
\usepackage{booktabs} 
\usepackage{longtable} 
\usepackage{caption}
\usepackage{subcaption}

\usepackage[english]{babel}
\renewcommand{\thesection}{\arabic{section}}
\renewcommand{\thesubsection}{\thesection.\arabic{subsection}}

\usepackage{enumitem}
\setlist{itemsep=0.3em, topsep=0.3em}
\usepackage{xurl}

\usepackage[runin]{abstract}

\usepackage{hyperref}

\hypersetup{
colorlinks=true,
urlcolor=blue,
linkcolor=blue,
citecolor=blue,
pdftitle={@title - @author},
pdfsubject={Systematic Review},
pdfauthor={@author},
pdfkeywords={Artificial Intelligence, Surveys and Overviews, AI use cases, Non-governmental organizations (NGOs), AI for
    Good, Technology Adoption, IT governance, Enterprise information systems}
}

\usepackage[backend=biber,style=numeric,sorting=none]{biblatex}
\usepackage{etoolbox}
\usepackage{indentfirst}

\usepackage{custom_commands} 

\usepackage{subfiles} 

\usepackage[pdftex]{graphicx} 

\begin{document}
\renewcommand{\abstractname}{} 

\maketitle 


\setlength{\absleftindent}{0em}
\setlength{\absrightindent}{0em}

\begin{abstract}
    \abstractText 
\end{abstract}

\section{Introduction}
\noindent
Artificial intelligence (AI) is quickly reshaping the way organizations gather information, make decisions, and provide services. This trend in adoption is reflected in the forecast for the global AI market, which is expected to reach \$1.81 trillion by 2030, with a projected compound annual growth rate (CAGR) of 35.9\% between 2025 and 2030 \cite{AI_Market_2025_2030}. While this widespread adoption in the private sector is getting increased attention, the parallel progression in the non-profit and charity sector has seen significantly less focus. Considering NGOs follow missions that increase social good, such as education, disaster relief, or democratization, it is in the public interest to ensure their work is as effective as possible.

For NGOs, AI has the potential to act as a catalyst, helping them improve efficiency, raise their public profile, and strengthen the reach of their initiatives. Among other applications, it can streamline routine tasks, offer data-driven guidance for program development, and support more focused and effective fundraising and advocacy efforts. For example, AI can be used to support NGOs in targeting donors more effectively and gaining insights into their likely contributions, potentially reducing campaign expenses and boosting fundraising outcomes \cite{madimi2023}. However, NGOs operate under a distinct set of constraints that affect their adoption journey. They often have to deal with tight budgets and an organizational culture that tends to be mission-centric rather than technology-driven \cite{godefroid2024identifying}. This creates a clear tension: Although NGOs face growing pressure to prove their effectiveness in the face of economic uncertainty and competition for funding, their unique financial and organizational limitations often prevent them from embracing the very technologies that could help them meet these demands. This tension between AI’s promised advantages and the practical limitations faced by NGOs represents a critical concern.

Although AI adoption in the non-profit sector is both highly important and shaped by distinctive dynamics, current research remains scattered and lacks a broad synthesis. 
Much of the existing research consists of isolated case studies or reports that either focus on a single country or a specific project. This can paint a biased and skewed picture of the overarching systematic dynamics of AI adoption in NGOs. While some systematic reviews in this field exist they either focus on narrow geographic contexts \cite{eng2024impact}, primarily emphasize ethical considerations \cite{schiff2021ai}, or concentrate on government and for-profit organizations \cite{tomazevic2024organizational, taboada2023artificial}. Therefore, a systematic review is needed that fills this research gap and synthesize the fragmented literature into a broad and nuanced summary that addresses how AI is adopted by NGOs. Specifically, there is a notable absence of systematic analysis that compares AI uses and barriers across different NGO sizes and geographic contexts, which would provide a more detailed and generalizable perspective.

To guide this research, we focused on two core research questions: \begin{itemize}
\item \textbf{RQ1}. In what ways is AI primarily utilized by NGOs and how does this differ based on size and location?
\item \textbf{RQ2}. What challenges and solutions are found in the literature about the adoption of AI in NGOs and how are these dynamics influenced by size and location?
\end{itemize}
In order to address these questions we draw on peer-reviewed studies as well as high-quality grey literature published in English between 2020 and 2025, ensuring a solid and up-to-date evidence base for the analysis.

This review makes three main contributions. For one it draws on the fragmented literature base to arrive at a comprehensive understanding of how AI is adopted and used by NGOs. Secondly, it investigates how AI adopting differs across different organizational contexts such as NGO size and geographic location. Finally, it also identifies emerging challenges and offers groundwork for practical guidance regarding NGOs’ highly individual journeys to technological adoption. The remainder of the paper is organized as follows: The next section outlines the theoretical background, followed by the methodology for conducting the systematic review, the results of the narrative and thematic synthesis, a discussion of the findings with implications and directions for future research, and finally, the conclusion.

\section{Theoretical Background}
For the purpose of this review AI is defined in a functional and applied sense, referring to computational systems designed to perform tasks that typically require human intelligence \cite{russell2020aima}. This encompasses but is not limited to tasks such as data analysis, predictive modelling, pattern recognition, and content generation through machine learning (ML) and large language models (LLMs). Throughout this paper the term NGO is used broadly for the non-profit sector, including international NGOs, local community-based organizations, and other civil society groups and charities. It acknowledges that this sector is not a singular homogeneous entity but encompasses a diverse set of mission-driven organizations that can vary significantly in their structure, resources, purpose and operational context \cite{marquez2016relevance}.

To provide a robust analytical structure for this review, the Technology-Organization-Environment (TOE) framework will serve as the primary conceptual lens.
This framework introduced by Tornatzky et al. \cite{tornatzky1990processes}  provides a powerful tool to analyse technology adaptation and implementation on an organizational-level based on three key contexts: The technology context describes what technical tools are currently used within the organization and are potentially available to it. Additionally, it incorporates the characteristics of the technology itself, including its complexity, perceived usefulness, and compatibility with existing systems. Secondly, the organizational context includes the internal characteristics of the organization, such as its size, management structure, resource availability, and culture. Thirdly, the environment context describes the external environment in which the organization operates, including donor expectations, the regulatory landscape, and local infrastructure.

While the framework is often viewed as generic and broad \cite{zhu2005post}, it still remains the gold-standard for analysis of technology adoption on an organizational level. Furthermore, it is particularly well-suited for this study because its three dimensions align directly with the research questions, allowing for a holistic and structured analysis of the challenges faced by NGOs in the context of organizational size and geographic context. In addition to TOE, the Diffusion of Innovations (DOI) theory \cite{rogers2003diffusion} is used as a secondary lens to understand the current state of AI adoption within the non-profit sector. DOI helps to explain why NGOs, as a social system, might be lacking behind in technology adoption compared to the private sector. It highlights that adoption follows an S-shaped curve influenced by the varying readiness of different adopter groups. It is commonly seen as being closely related to TOE and thus methodologically fit for this review \cite{baker2011technology}.

This systematic review will utilize a combination of narrative and thematic synthesis approaches, which is particularly well suited given the heterogeneous nature of the available literature, which includes both quantitative survey data and qualitative case studies. This data heterogeneity also makes a meta-analysis unfit in this context. The narrative synthesis approach goes beyond a simple summary but tells a structured and thematic story of the literature, that allows for identifying patterns and integrating diverse types of evidence into a coherent understanding of the topic. This method will use the TOE and DOI frameworks as scaffolding for guidance and will be thematically oriented at the research question. This approach ensures that the findings are well organized and rigorous.

Given the limited structured syntheses on AI adoption within NGOs, especially when compared to the private sector, this study seeks to address that gap and offer a cohesive understanding of the topic. 

\section{Methodology}
\begin{figure*}[h!] 
    \centering
    \includegraphics[width=\textwidth]{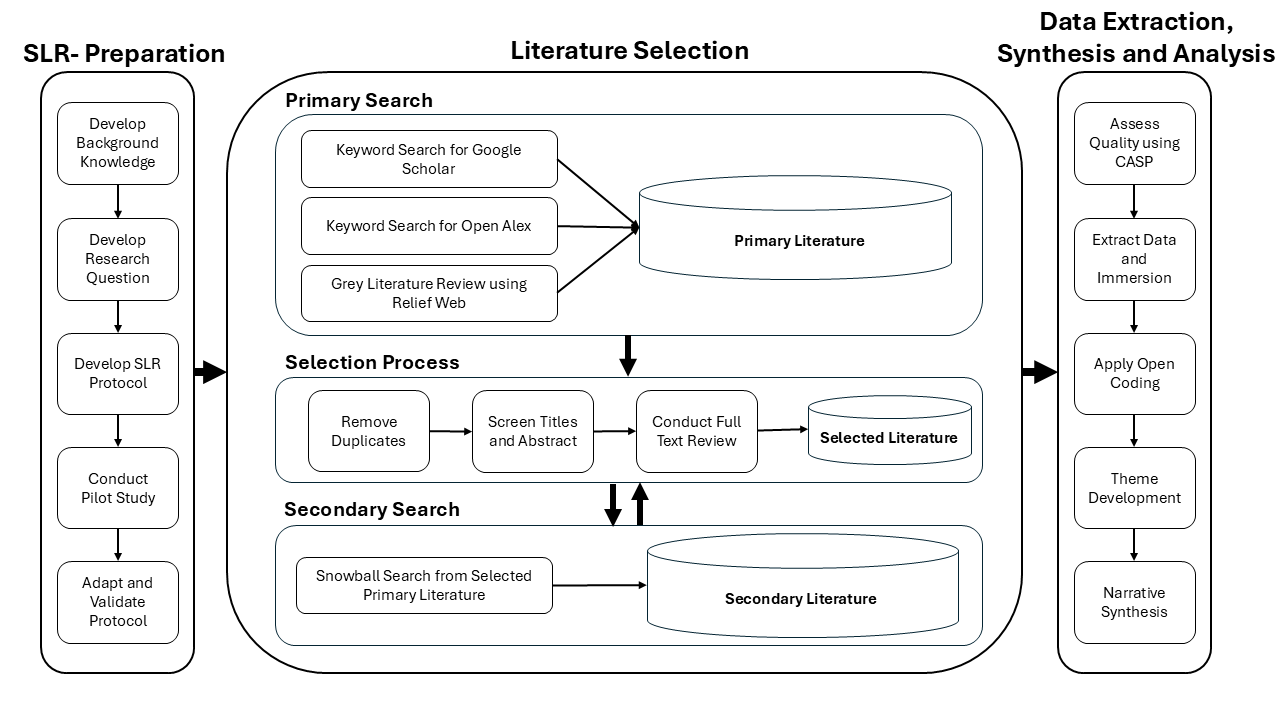} 
    \caption{Pipeline for this SLR. Note that this graphic is inspired by \cite{shams2025ai}}
    \label{fig:pipeline}
\end{figure*}

This study explores and aims to gain a comprehensive understanding of the adoption and usage of AI by a diverse set of NGOs on a global scale from the published research literature. Our
research was guided by the two research questions defined in the introduction.
We conducted a systematic literature review (SLR) following the PRISMA 2020 guidelines proposed by Page et al. \cite{page2021prisma}. This framework is widely adopted in contemporary literature and is generally regarded as best practice for conducting SLRs. It offers a comprehensive checklist to guide the design, execution, and reporting of systematic reviews, ensuring rigour and transparency throughout the research process.

Additionally, we utilized the SPIDER framework developed by Cooke et al. \cite{cooke2012beyond}. This search strategy tool improves over the PICO paradigm \cite{richardson1995well} by offering improved efficiency for researchers, thereby saving them valuable time by eliminating irrelevant articles. It is additionally developed to particularly fit qualitative and mixed-methods research, something highly desirable in the context of this SLR.

The components of the SPIDER framework in this review are the following: 
\begin{itemize}
\item \textbf{S}ample: Non-Governmental Organizations (NGOs), Non-Profit Organizations (NPOs), and international aid agencies. 
\item \textbf{P}henomenon of \textbf{I}nterest: The utilization and application of Artificial Intelligence
\item \textbf{D}esign: Empirical studies, case studies, and technical reports documenting AI applications
\item \textbf{E}valuation: The evaluation will focus on the social impact, efficiency gains, and operational improvements, as well as the associated ethical and practical considerations.
\item \textbf{R}esearch Type: Qualitative and quantitative research, observational studies, and case studies, recognizing that a randomized controlled trial design is highly unlikely in this field.
\end{itemize}

Before conducting the full systematic search, we carried out a pilot study to test and refine our search strategy and eligibility criteria. 
The initial search queries were run in Google Scholar and OpenAlex, and the first 40 results from each database were screened against the preliminary eligibility criteria. 
This step revealed that while the core concepts of "artificial intelligence" and "NGOs" retrieved relevant literature, several irrelevant results were also captured, 
particularly in domains unrelated to the social sector (e.g. AI for business analytics or government applications). 
Based on insights from the pilot study, the search strings were iteratively refined in several ways. First, additional synonyms and related concepts (e.g. charity, NLP) were incorporated to broaden coverage. Second, certain terms were restricted to the title or abstract fields to increase precision. In particular, the term 'NGO' was limited to the title and abstract field, since 'Ngo' is also a common Vietnamese surname that could otherwise introduce irrelevant results. Third, the use of wildcards was applied more selectively to increase retrieval accuracy and reduce irrelevant matches. Fourth, the set of application-related terms was expanded by including additional fields synonymous with social impact. Finally, the eligibility criteria regarding population and context were slightly clarified to ensure consistency and alignment with the objectives of this review.

After refinement, the pilot search produced a substantially higher proportion of relevant hits (approximately 49\% vs. 23\% in the first iteration) as assessed by title and abstract, 
which confirmed that the strategy was sufficiently sensitive and specific to be used for the main review. 
The initial and finalized search strings are documented in Appendix~\ref{appendix:search querys}. Generally they are derived from the Sample, Phenomenon of Interest and Evaluation definitions given in the SPIDER framework. Please note that Design and Research Type are dealt with by the eligibility criteria.

\subsection{Primary Search}

In accordance to the findings by Martin et al. \cite{martin2021google} at least two databases were used for the primary search. Due to the relative novelty of the field the two databases with the highest coverage, Google Scholar and OpenAlex \cite{walters2025comparing} were chosen. Additionally, Scopus was used. 

To mitigate publication bias, a thorough search of grey literature was conducted \cite{adams2017shades}. This included technical reports, white papers from NGOs, and government reports that may not be indexed in academic databases. These documents were collected through Relief Web, a repository featuring official UN and NGO reports. All of the findings are included or excluded according to the same criteria as the white literature.

In the context of this SLR the eligibility criteria were the following: 
\begin{itemize}
\item \textbf{Population}: The study must involve a Non-Governmental Organization (NGO), Non-Profit Organization (NPO), charity, or a similar civil society organization. Studies focused on for-profit companies or government agencies including UN bodies are excluded.  

\item \textbf{Intervention}: The study must detail the use or application of AI, machine learning (ML), natural language processing (NLP), or other closely related AI technologies.

\item \textbf{Context}: Studies must be related to the social impact sector. This could include humanitarian aid, social services, environmental conservation, public health, or human rights.

\item \textbf{Study Design}: The review will utilize primary research, such as case studies, empirical studies, or project reports. Secondary publications like editorials, commentaries, and other literature reviews are ignored to avoid redundant information and potential bias. 

\item \textbf{Research Type}: Included studies are qualitative as well as quantitative research. Master theses and PhD Dissertations are excluded to ensure a minimum quality standard and mitigate the issue of accessibility and consistency between different databases.

\item \textbf{Publication Dates \& Language}: The review investigates the time frame from the emergence of modern AI to the present to capture the state-of-the-art in this rapidly evolving field (2020–2025). The review is limited to publications in English, as this is the only languages in which the contributing authors all have sufficient proficiency in to accurately interpret, assess, and synthesize the content. Including other languages could compromise the reliability of data extraction and analysis.

\item \textbf{Practical Considerations}: Literature is excluded if the full paper is unavailable or not freely available online under Universitat Pomepu Fabra literature agreements.
\end{itemize}

All of these criteria were evaluated and slightly refined as part of the pilot study to ensure relevance and consistency. For an exact overview of the number of included studies  and reasons for exclusion, please refer to Figure \ref{fig:flowchart}.

\subsection{Selection Process}
Following the recommendations of Pigott et al. \cite{pigott2020methodological}, the screening process was carried out in sequential steps:
\begin{enumerate}
\item Duplicate citations from different databases were removed.
\item The titles and abstracts were screened for relevance by two independent researchers to reduce bias and increase reliability and consistency. Reviewers were not blinded to the journal or author information. Any discrepancies in the screening were resolved through discussion until a consensus was reached. If a consensus could not be reached, a third person (a mutually agreed-upon arbitrator) made the final decision. Only studies that clearly did not meet one or more of the eligibility criteria were removed.
\item The remaining articles underwent full-text screening. In this stage remaining literature was reviewed in detail against the full set of eligibility criteria to ensure the study design, population, and intervention align with this review's scope.
\end{enumerate}

Although the review was conducted by only two authors, which could introduce a slight risk of bias, several measures were implemented to minimize this risk. Both authors worked independently during study selection, data extraction, and coding, and methodological rigour was maintained by adhering to a pre-defined review protocol found under this \href{https://osf.io/a2357/?view_only=7120b178719540608dc97cd2a178ceff}{Open Science Framework (OSF) link}. Regular discussions and cross-checking ensured consistency and transparency, helping to mitigate potential biases and strengthen the reliability of the findings. Inter-annotator reliability was assessed using Cohen’s Kappa. Agreement was substantial for Google Scholar ($\kappa$ = 0.76), Open Alex ($\kappa$ = 0.70) and Scopus ($\kappa$ = 0.72) and perfect for ReliefWeb ($\kappa$  = 1.00). Overall Cohen's Kappa was also substantial ($\kappa$ = 0.74).

\subsection{Secondary Search}

\begin{figure*}[h!]
    \centering
    \includegraphics[width=\textwidth]{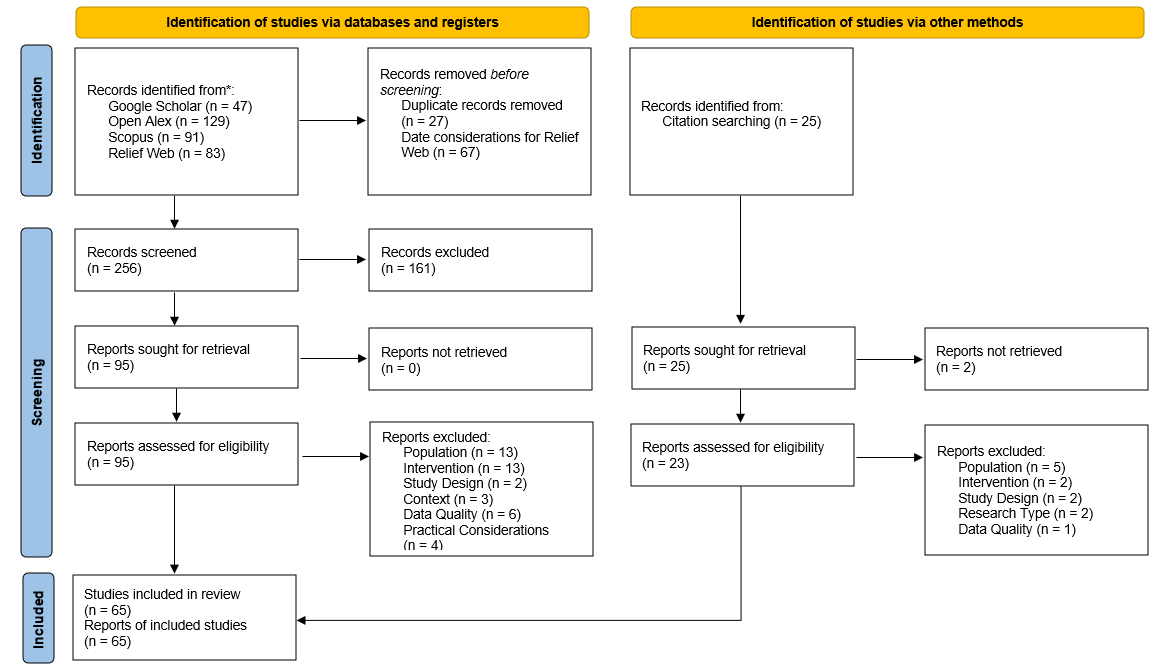} 
    \caption{PRISMA flowchart as defined by Page et al.\cite{page2021prisma}}
    \label{fig:flowchart}
\end{figure*}

After identifying a set of key studies in the primary search using the selection process, we employed forward and backward citation tracking (snowballing) to increase the body of relevant literature \cite{hirt2023citation}. Snowballing can be beneficial as it helps capture relevant studies that may not appear in database searches due to indexing limitations or differing terminology. Initially, studies found this way were judged based on their title and abstract. Later on, a full review of remaining studies was conducted to ensure they meet the eligibility criteria. An overview of the exact number of studies identified through this process and potential reasons for their exclusion is provided in Figure \ref{fig:flowchart}.

\subsection{Quality Assessment}
To ensure only high quality studies are used in this review the CASP (Critical Appraisal
Skills Programme) paradigm \cite{CASP2018} was deployed. This quality assessment tool
includes multiple checklists suitable for qualitative studies, case studies, mixed-method research and quantitative studies and provides
clear questions to assess rigor, credibility, and relevance. Depending on the methodology used in the study it consists of between ten and twelve questions (evaluating validity, results and the value of the research) which can be answered either with "Yes", "No" or "Can't tell". As the review included a number of quantitative and algorithmic studies, the CASP checklists were adapted to better reflect aspects of quantitative rigor, such as statistical validity, reproducibility, data transparency, and performance evaluation. In accordance to common practice we considered studies which have two or more questions answered with "No" as too poor quality to take them into account for the SLR.

Of the studies remaining after the selection process and the secondary search, 43 did not receive any “No” responses. 16 studies received a single “No” response, while seven received two or more and were thus excluded. The remaining six documents were white papers, which are not fit for appraisal using the CASP framework. These were instead evaluated for internal validity and the credibility of their sources, both of which were all documents deemed sufficient.

\subsection{Data Extraction}
Data from the included studies were extracted using a structured Excel spreadsheet to systematically capture both demographic and content-related information. Demographic data included bibliographic details such as author(s), year of publication, title, journal or source, and geographic location of the study. Study characteristics such as study design (e.g. case study, survey), NGO context (e.g. size, mission), and AI application specifics (e.g. type of AI, function, and specific application) were also recorded. Content-related data included reported findings and outcomes (quantitative results and qualitative observations), ethical considerations (e.g. bias, data privacy, transparency, accountability), and limitations highlighted in the original studies.

Data extraction was conducted manually by the first author and cross-checked with the second authors to ensure consistency and accuracy. A structured extraction form was piloted on a subset of studies prior to full extraction to confirm that all relevant information could be reliably captured.

\subsection{Data synthesis and analysis}
Due to the heterogeneity of the included studies and the limited availability of directly comparable quantitative data, a meta-analysis was deemed inappropriate. Instead, a qualitative approach combining narrative and thematic synthesis was employed to interpret and summarize the findings, providing a nuanced and coherent overview of the field \cite{mohammed2016meta}. Thematic synthesis involved several steps: reviewers first immersed themselves in the studies, reading them thoroughly to gain a comprehensive understanding of the data. Key words, phrases, and concepts were then systematically and manually coded using both deductive codes (derived from the research questions, e.g. “AI use in donor management”) and inductive codes (emerging directly from the data, e.g. a specific use case of AI). Related codes were subsequently manually grouped into broader, overarching themes, with thematic mapping used as a visual aid to identify relationships between themes. A simplified version of the final thematic mapping can be found in appendix \ref{app:thematic map} and a full overview is available in the \href{https://osf.io/a2357/?view_only=7120b178719540608dc97cd2a178ceff}{additional material}. Finally, these themes were integrated into a cohesive narrative that synthesized the evidence and directly addressed the research questions. While the process was primarily carried out by the first author, who repeatedly re-read the literature to gain deeper immersion, the second author was actively involved throughout the review. The final results were determined through multiple rounds of iterative discussions among all authors, leading to a consensus on the completed mapping.

\section{Results}
This section reports the findings of the systematic literature review, beginning with an overview of the characteristics of the 65 included studies. Furthermore, we analyse the extracted relevant findings regarding the different utilization of AI by NGOs as well as the emerging challenges and solution regarding the implementation process. We specifically focus on how this differs based on location and size of the NGO.

For operationalization purposes, we decided to categorize the studies regarding location and size. This is beneficial for later thematic and structured analysis of relevant findings. For the location component we drew on the World Bank country classifications by income level for 2024-2025 \cite{metreau2024world} which classifies each country based on the Gross National Income (GNI) per capita in four income levels: Low income, lower-middle income, higher-middle income and high income. This allows to move beyond the simplistic definition of developing vs. developed countries towards a more nuanced understanding of the economic situation. This is particularly important in the context of this SLR, as differences in infrastructure and access to technology and resources can influence the adoption of AI in NGOs. Furthermore, this classification enables more meaningful comparisons across countries with similar economic conditions.

We also classified NGOs by size into three different categories based on their number of employees, as assessed by publicly available information like corresponding websites: Small (less than 15 employees), moderate (between 15 and 50 employees) and large (more than 50 employees). 

\subsection{Characteristics of included studies}

\begin{figure*}[h!]
    \centering
    \begin{subfigure}[b]{0.48\textwidth}
        \centering
        \includegraphics[width=\textwidth]{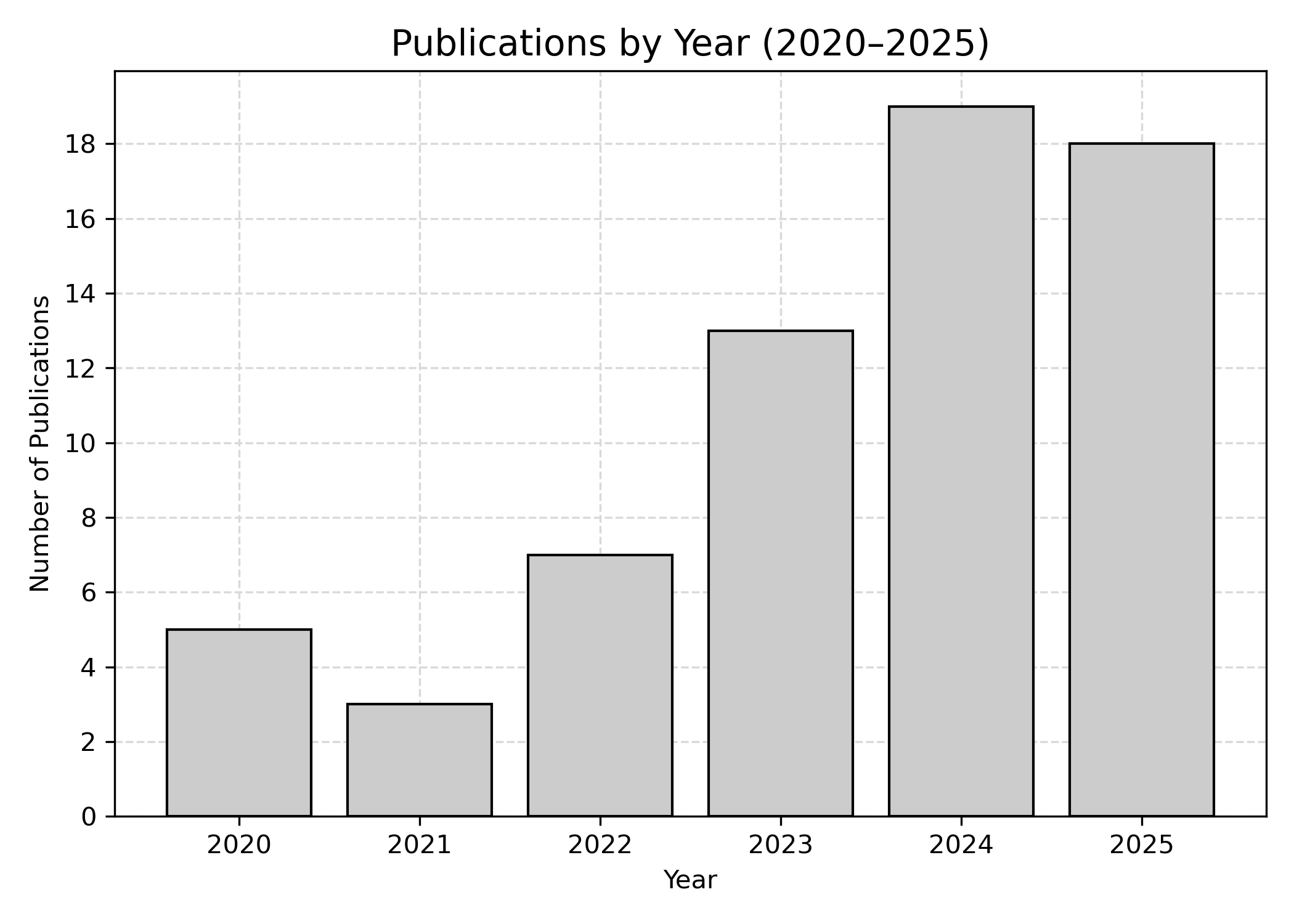}
        \caption{Publications by Year (2020–2025)}
        \label{fig:bar_chart}
    \end{subfigure}
    \hfill
    \begin{subfigure}[b]{0.48\textwidth}
        \centering
        \includegraphics[width=\textwidth]{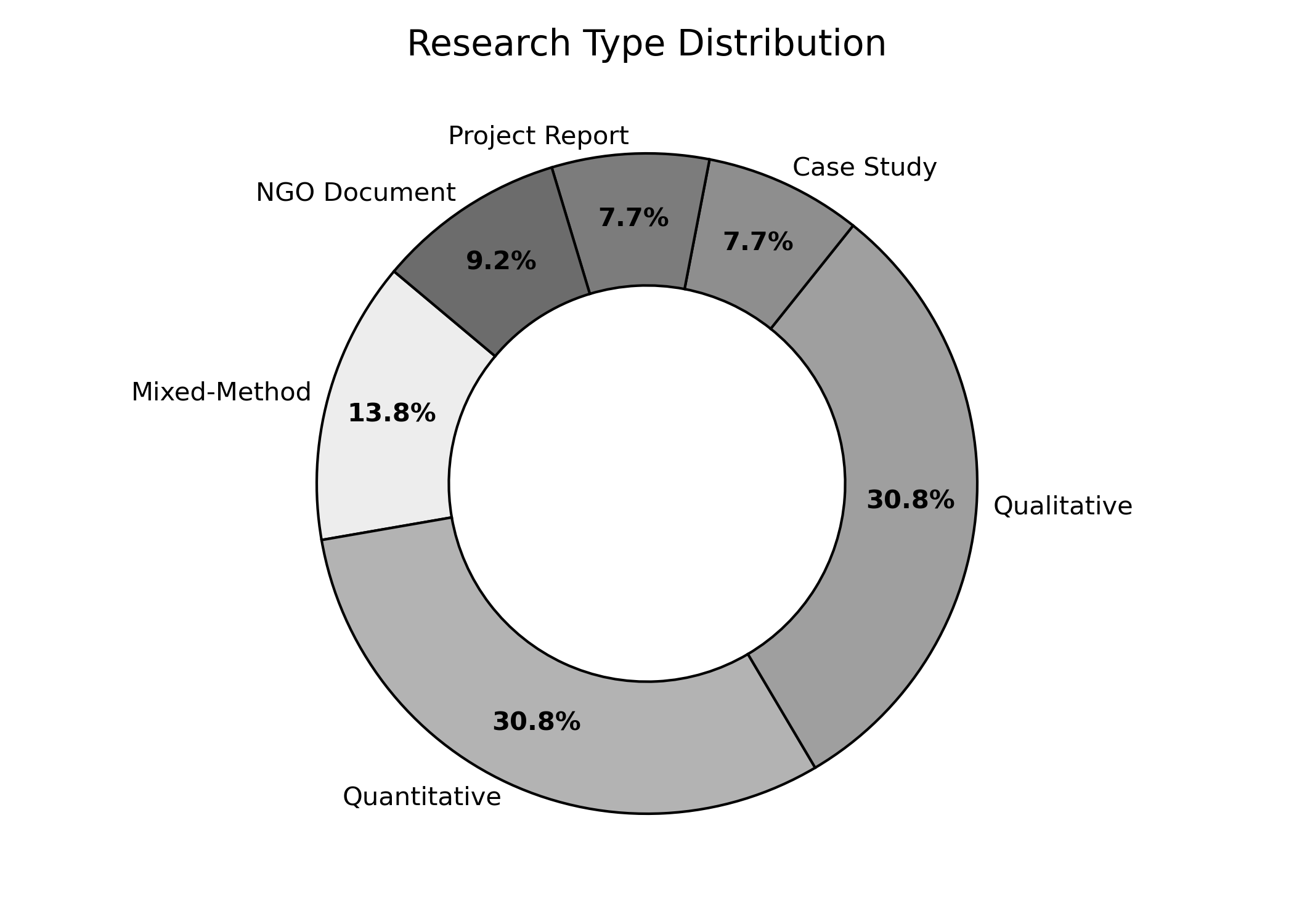}
        \caption{Research Type Distribution}
        \label{fig:donut_chart}
    \end{subfigure}
    \caption{Visualization of publication dates and research methodology}
    \label{fig:demographic}
\end{figure*}

As illustrated in Figure \ref{fig:demographic} the number of studies on AI adoption in NGOs has shown a steady increase since 2021 (considering only around three quaters of the year 2025 have already passed at the time of writing), highlighting the growing importance of the research area. Furthermore, included studies had an approximate balanced research type between quantitative and qualitative methods. This exemplifies the variety of approaches to this complex topic and improves the reliability of this SLR, as it draws on findings derived with multiple different approaches. Moreover, the studies were published in a variety of journals and conferences, further emphasizing the breadth of scholarly interest. For an tabular overview of included studies and their discussed topics please refer to appendix \ref{app:overview}. An even more detailed overview of a matching between journals/conferences, key findings, research methodology, ethical considerations and limitations to the corresponding papers, is provided in the \href{https://osf.io/a2357/?view_only=7120b178719540608dc97cd2a178ceff}{additional material}.

Unsurprisingly, most research was dedicated to large NGOs (29.3\%). This was followed by research specifically concerning moderate (12.3\%) and small (9.2\%) NGOs. The other studies (53.8\%) were unspecific on which type of NGO they concerned. Regarding the location of NGOs, most of the included studies pertained to high (26.2\%) or lower-middle (26.2\%) income countries, whereas only a minority concerned medium-high (12.3\%) and lower income (3.1\%) nations. The rest of the studies (46.2\%) concerned either NGOs that acted globally or were unspecific to the location of NGOs they concerned. Please note that these numbers do not necessarily sum up to 100\% as some studies concerned multiple NGOs of different sizes and with different geographic contexts.

This demonstrates a significant gap in the literature, as most available research focuses on large NGOs and higher or lower-middle income nations. However, AI adoption in small NGOs or NGOs operating within low income countries remains largely unexplored. Possible explanations for this biased focus include structural barriers, as small NGOs tend to have less capacity to cooperate with researchers and lower income nations typically lack digital infrastructure required for academic collaborations. Furthermore, data availability concerns might play a role, as research from low-income nations is often less visible and accessible in academic databases. Another possible explanation might be a bias in research interest.

Another notable characteristic is that among the 31 studies developing new AI innovations for NGOs, only about half actively involved NGOs as collaborators (32.3\%) or had the NGO driving the development process (12.9\%). The remaining studies either treated NGOs merely as data providers and end users (9.7\%) or designed solutions without any active NGO engagement (45.1\%). This limited engagement suggests a potential misalignment between AI innovations and NGO needs, which may hinder adoption, reduce practical impact, and limit the contextual relevance of these tools.

\subsection{Utilization of AI in NGOs (RQ1)}
\begin{figure*}[h!]
    \centering
    \includegraphics[width=\textwidth]{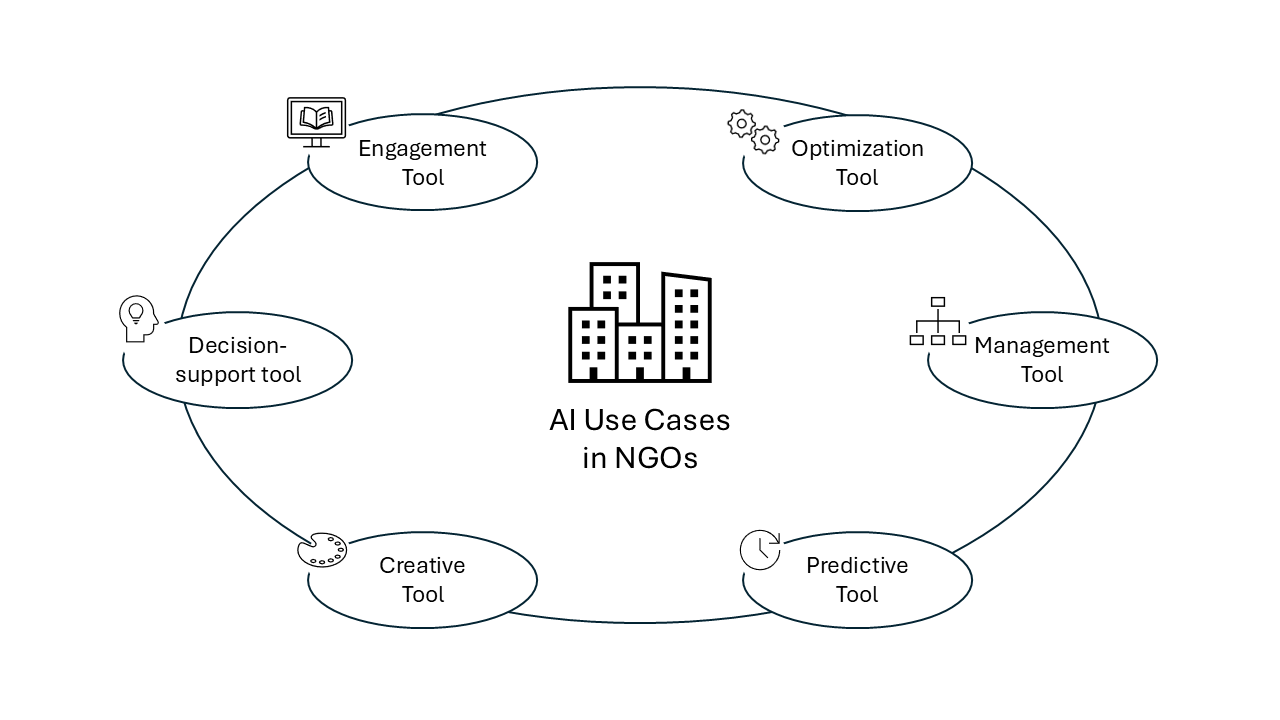} 
    \caption{Overview of identified use cases of AI in NGOs}
    \label{fig:usecases}
\end{figure*}

The adoption of AI in NGOs is still in its infancy. Several studies report that NGOs have been slow to leverage AI \cite{dubey2022impact, atalay2025ai, ahatsi2025humanitarian, scutto2025empowerment} and even when they do, the perceived impact remains limited \cite{nejc2025role}. Depending on sector geographic context and size, research indicates that between 7.6\% \cite{kazemeini2025dataset} and 40\% \cite{dube2024factors} of NGOs utilize AI in some shape for now. Drawing on the DOI model, this positions AI adoption in NGOs between the early adopter and early majority stage \cite{rogers2003diffusion}. This stage is characterized by cautious but growing experimentation, visible success stories that begin to influence peers, and uneven diffusion shaped by resources and infrastructure. At present, AI serves primarily as a supportive resource rather than a replacement for human judgment, with some larger projects being implemented and a highly variable degree of organizational integration across the sector \cite{hahn2025strategic}. In the following, we identify and six primary categories of how AI is utilized by NGOs that are visualized in figure \ref{fig:usecases}.

One prominent method NGOs utilize AI is as an \textit{engagement tool}. In one study amongst a variety of non-profits in size, speciality, and area, 60\% of participants reported that AI-driven engagement solutions increased stakeholder interaction and lead to better satisfaction scores \cite{faruq2024ai}. In the educational sector, for instance, AI can be utilized by NGOs to improve educational outcomes through individualization via personalized learning platforms that can be deployed in underserved communities where access to quality education is limited \cite{ghani2021ml, elamin2024ai}. Early adopters like the International Rescue Committee \cite{toplic2020ai}, Khan Academy \cite{unesco2023sal} or a portion of NGO owned schools in Egypt already use and investigate its application \cite{soudi2023generative}.

Another promising strand of AI-supported engagement is its application in interactive chatbots. Large NGOs employ these systems to provide refugees with accessible, localized information about their rights \cite{efthymiou2023role, weber2024emergency, toplic2020ai} and to enhance the communication with beneficiaries through round-the-clock availability \cite{cheng2025aifeed}. Chatbots expend organizational reach and engagement through interaction with a larger audience simultaneously, saving costs in the long run, and freeing up time for staff \cite{dube2024factors}.

A further significant use case is the application of AI as an \textit{optimization tool}. In the healthcare domain, AI supports critical innovations such as optimizing vaccine distribution in lower and lower-middle income countries \cite{nair2022adviser}, effectively delivering timely preventive care information to new and expecting mothers \cite{mate2022bandits}, and reducing total response time in emergency medical services \cite{rathore2022sustainable}.

Furthermore, AI as an optimization tool is also used in service delivery by organizations like the Danish Refugee Council or the International Rescue Committee that leverage it to optimize their routing and enhance timely delivery to refugees \cite{efthymiou2023role, toplic2020ai}. Similarly, another study developed an AI-powered tool for optimizing food delivery in cooperation with Ekram, a small Saudi Arabia NGO, and demonstrated increased effectiveness in the delivery process \cite{alhindi2020vehicle}. A further use case lies in the application in disaster relief, where AI and big data analytics capabilities (AI-BDAC) were shown to significantly increase resilience and humanitarian supply chain agility in NGOs \cite{efthymiou2023role, dubey2022impact, pereira2024ai}. 

Beyond optimized service delivery, the optimization capabilities of AI also extend to sustainability efforts. By helping to develop effective long-term strategies for protected areas \cite{atalay2025ai} or optimize biodegradable plastic packaging design \cite{ulfy2025ai} AI was also shown to effectively enhance green initiatives. 

The most prominent utilization of AI as an optimization tool, however, remains its application in donation management. Here AI enhances donor matching with beneficiaries \cite{kapuge2024optimizing}, keeps track and secures physical donations \cite{pai2023ngo, laylo2023impact}, helps with verification of eligible recipients \cite{laylo2023impact}, and analyses donor behaviour for more effective fundraising \cite{efthymiou2023role}. Data taken from a wide range of NGOs with humanitarian missions through quantitative surveys revealed that total funds raised increased by an average 20\% when using AI applications \cite{faruq2024ai}. However, some researchers note that content and objectives of fundraising strategies are unlikely to change as a result of AI. Nonetheless, we might see how increasing automation may shift strategies from broad target groups to fully individualized approaches \cite{hahn2025strategic}.

NGOs, particularly those in disaster relief, often face challenges that require fast and well-informed decision-making. AI as a \textit{decision-support tool} is one solution to address this issue. In public health, for instance, AI has been deployed to forecast disease outbreaks \cite{elamin2024ai}, identify high-risk populations for targeted interventions like during the Ebola pandemic in West Africa \cite{ghani2021ml}, predict air quality in India \cite{ghani2021ml} or analyse food quality for charitable donations \cite{uke2024food}. Furthermore, AI-powered decision-support tools can be leveraged to inform project funding and where to deploy resources \cite{amalraj2025ml} or assist in the development of more effective intervention strategies for sextortion \cite{alex2023designing} or cyber bullying cases \cite{gupta2024cyber}. In Disaster Relief, AI can help by providing real-time information for emergency response for example via social media analysis \cite{efthymiou2023role, weber2024emergency}. For example, Care Centre, a large global NGO, used AI to get more accurate and timely analysis of the Syrian conflict \cite{toplic2020ai}.

For NGOs, another promising path forward is adopting AI as a \textit{management tool}. For instance, AI functions as a powerful instrument for optimizing resource allocation and mitigating risks related to financial instability \cite{huang2024tech, ghani2021ml}, while increasing efficiency and reducing errors in accounting and procurement \cite{krause2025ai, bahameish2023ai}. A study amongst global NGO representatives showed that AI usage reduced work hours needed for data entry by 60\% and for resource allocation by 30\%  \cite{faruq2024ai}. AI has also been employed to predict the financial requirements of various charitable projects \cite{sanjana2024integrating} and for evaluating program impact by analysing outcome data \cite{efthymiou2023role}. Supporting this, Alnamrouti et al. demonstrated that AI also has a positive and significant impact on sustainable organizational performance \cite{alnamrouti2022ai}. 

In staff management AI-powered tools can be utilized to screen and hire promising employees \cite{efthymiou2023role} or volunteers \cite{avagyan2020utilizing}, while reducing costs and solving recruiting delays \cite{asajile2024ai}. However, the main agency for human resource management still remains largely human, with technology adoption being minimal and serving rudimentary or supporting functionalities \cite{dutta2024machine}. Another crucial part of NGOs day-to-day work is the integration of volunteers. This area has also benefited from AI-powered platforms such as IDEA NGO \cite{kannan2025ai} or UniteVol \cite{bhuvaneswari2025unite}, which facilitate efficient matching between organizations and prospective volunteers \cite{sharma2021ocr}.

AI as a management tool can also simplify internal organizational flow by automating routine conversations and improving event planning \cite{efthymiou2023role, joshi2025smart, popescu2024impact}. Its integration has been identified as a significant predictor and moderator of long-term organizational performance \cite{alnamrouti2022ai}. However, adoption is not without drawbacks: One study found that the use of AI increased the time required for internal communication by 25\% among global NGOs \cite{faruq2024ai}.

Another promising area for NGOs lies in the area of leveraging AI as a \textit{creative tool}. Among charity workers in the UK, AI is increasingly recognized for its potential to improve efficiencies in content generation \cite{hansen2025ai}, and even conservative organizations like churches see potential for its utilization in areas such as sermon preparation \cite{sievert2024ai}. In fact, NGOs already use AI to help write and refine individualized donor thank-you notes, newsletters, grant proposals, and press releases \cite{sandberg2025ai, popescu2024impact}.

Beyond communication and fundraising, AI also supports educational and awareness-raising initiatives. Examples include the assistance in preparing educational material for using food in healthy ways via recipe generation \cite{sammer2024aifeed} or the refinement of resources for informing about sextortion \cite{alex2023designing}. More experimental approaches emphasize participatory creativity. For instance Crea.vision, a platform following a citizen-science approach focused on human-AI collaboration, allows users to create images on societal issues utilizing AI, and being subsequently linked to relevant NGOs \cite{rafner2023crea}.

AI as a \textit{predictive tool} can also assist NGOs in diverse ways. For instance, local NGOs have used AI models based on satellite data to predict shelter locations of refugees in Syria \cite{grass2023ml}, Columbia \cite{tingzon2020mapping}, or West Africa \cite{toplic2020ai}. Another example is the EUMigraTool (EMT), a joint European initiative using AI to predict short- and mid-term migration. EMT assists NGOs by guiding strategy development, informing funding decisions, supporting advocacy, and addressing gender-specific needs such as safe spaces and reproductive healthcare \cite{blasi2024eumigratool}. Similarly, the Danish Refugee Council (DRC) \cite{kbah2023understanding} and Save The Children \cite{scutto2025empowerment}, both large NGOs, developed separate forecasting models for predicting forced displacement that informs plan development and policy formulation.

Predictive modelling also play a key role in disaster preparation. Applications range from anticipating of conflicts and crises \cite{toplic2020ai}, to forecasting of climate change impact on vulnerable groups \cite{ghani2021ml} and predicting natural disasters \cite{elamin2024ai, weber2024emergency, ahatsi2025resilience, rahman2021development}. At a more localized scale, a study in Texas developed a Deep Neural Network that helps predict tenants-at-risk of eviction in cooperation with Child Poverty Action Lab, a  moderate sized NGO \cite{tabar2022forecasting}.

Overall, we identified six different categories of utilizations of AI in NGOs. However, the exact degree of application and usage differs with the size and geographical context of the organization. Large organizations like the red cross are reported to often utilize AI for more niche subjects like data visualization \cite{kravchuk2025ethical}. Due to their comparatively higher financial budget and their ability to hire domain experts \cite{niranjana2023ai}, they can also develop in-house solutions in cooperation with academic and industry partners. For example, Mercy Corps use an early warning system they developed in partnership with IBM \cite{efthymiou2023role}. According to the literature body, the usage of AI as an engagement, predictive and optimization tool is primarily limited to larger NGOs, while smaller NGOs already adopt it as a management and creative solution. 

The location of the corresponding NGO also plays a crucial role in the type of use cases of AI. High income countries which have advanced technological infrastructure posses a number of NGOs that are extremely technology ready and specifically focus on integrating technological innovations. For example, the NGO STEM for Dance in the US build an AI-powered virtual coding environment to code interactive dance routines in cooperation with academic partners \cite{castro2022ai}. Another example is Qatar Foundation (QF), which developed a comprehensive multi-year strategy on the integration of AI \cite{bahameish2023ai}. On the other hand, different sociocultural challenges also hold new room for AI applications. For instance, AI can be used by NGOs to help streamline the distribution process of Zakat, a Muslim tradition of charitable giving \cite{laylo2023impact, defnizal2020ai}. Unlike high income countries, many low- and lower-middle income nations, like India, have decentralized emergency medical services involving both semi-government and non-government organizations, which is another area of application of AI for NGOs \cite{rathore2022sustainable}. 

\subsection{Challenges and Solutions in AI Adoption (RQ2)}
\begin{figure*}[h!]
    \centering
    \includegraphics[width=\textwidth]{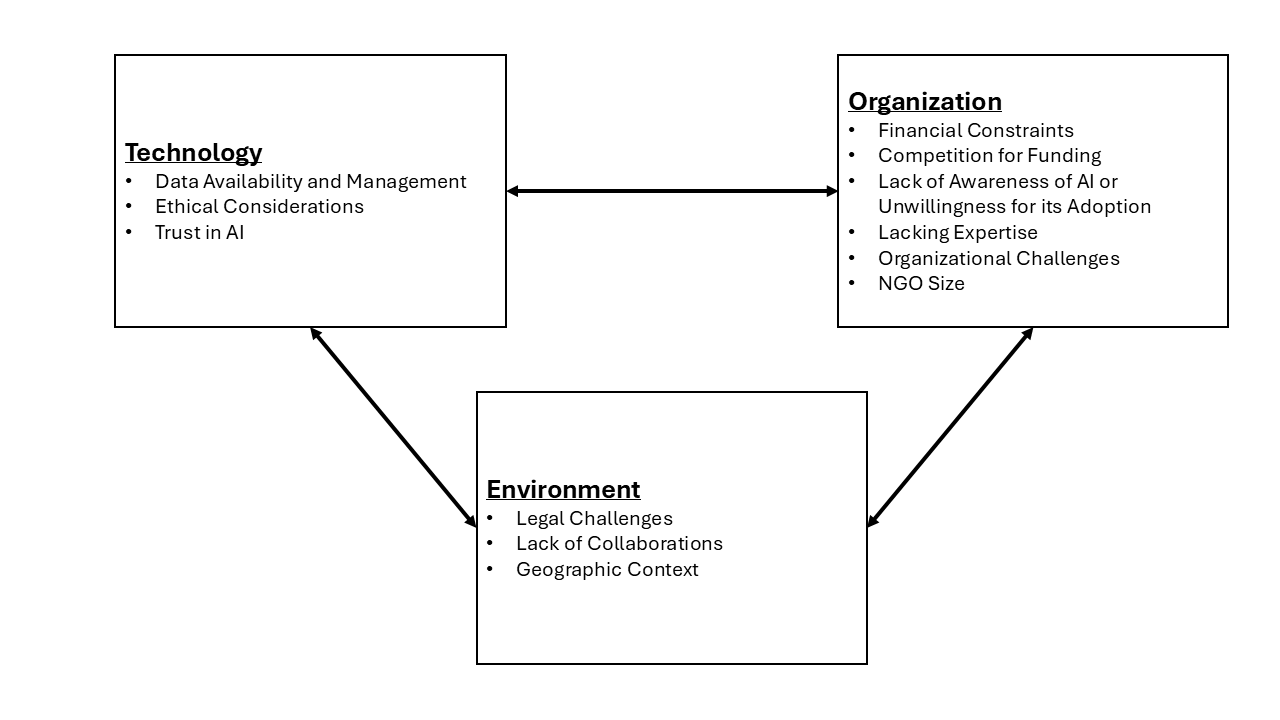} 
    \caption{Summary of identified challenges}
    \label{fig:challenges}
\end{figure*}

The adoption of AI by NGOs faces numerous challenges and obstacles. The TOE framework provides a useful lens for categorizing and understanding these barriers \cite{tornatzky1990processes}. Figure \ref{fig:challenges} gives an overview of the identified challenges, matched to the corresponding parts of the TOE framework.

\subsubsection{Challenges in Technology Context}

The technology context centres around the internal and external technologies available to an organization itself. For instance, AI at its core always requires some form of data to work with. \textit{Data availability and management} are thus a central concern, which can negatively affect the accuracy of machine learning predictions \cite{efthymiou2023role, ahatsi2025humanitarian}. NGOs are often faced with complex data that includes multiple possibly delayed data streams, diverse data structures, multiple ways of data acquisition \cite{nethope2024empowering, kbah2023understanding} and incomplete records \cite{ghani2021ml}, particularly in lower income regions \cite{grass2023ml}. Therefore, establishing robust internal data governance becomes a key challenge in the AI adoption journey \cite{elamin2024ai}. From an academic perspective, researchers such as Mate et al. argue that these challenges should be viewed as genuine research question rather than limitations \cite{mate2022bandits}. However, other research finds that in specific scenarios, like the DRC forecasting project, the high availability of data can also be an enabler for AI integration \cite{kbah2023understanding}.

Another key concern within the technological context involves \textit{ethical considerations}. NGOs must ensure that the adoption of AI not only aligns with their unique missions and goals \cite{pai2023ngo} but also complies with broader ethical standards. A central issue is the fairness and accountability of AI solutions \cite{nejc2025role}. Since these systems heavily rely on the quality of data, flawed or incomplete datasets can lead to biased outcomes that systematically disadvantage already marginalized groups \cite{ghani2021ml, kravchuk2025ethical, hansen2025ai, gupta2024cyber}. Especially in the midst of an emergency, humanitarian staff might not have the time to verify the information provided by the AI solution \cite{weber2024emergency}. This issue relates back to the opacity of algorithms, underscoring the importance of human-in-the-loop approaches \cite{kravchuk2025ethical, hansen2025ai}, particularly in areas that rely on individualized outputs like education \cite{soudi2023generative}. 

Beyond fairness and accountability, the risk of overreliance on technology represents another critical concern. NGOs tend to be cautious about ethical matters, as they do not want to over rely on technological solutions, emphasizing their mission-driven orientation in contrast to becoming technology-driven \cite{nejc2025role}. Safeguarding the ability to remain functional, even without the usage of AI, is a key concern regarding the critical independence of NGOs \cite{amalraj2025ml, weber2024emergency, popescu2024impact}.

Another key dimension of ethical considerations concerns security and privacy. NGOs need to comply with data confidentiality and ensure that data is stored securely \cite{kravchuk2025ethical, dube2024factors, faruq2024ai, unesco2023sal, alex2023designing, krause2025ai, iazzolino2025trading, gupta2024cyber}. Data servers are also frequently located abroad, outside the direct control of NGOs, increasing data security concerns \cite{hahn2025strategic}. For NGOs in particular, breaches of data security carry severe consequences, as they can result in significant losses of trust within the communities they serve \cite{ghani2021ml}. Overall, research emphasizes the need for NGOs to strike a balance between data protection and program accessibility \cite{kravchuk2025ethical}.

Beyond classical approaches to ethics, Sandberg et al. \cite{sandberg2025ai} developed a data feminist pedagogy regarding AI in NGOs, framing AI not merely as a technological tool but as a political instrument that reflects and reinforces prevailing values. From this perspective, NGOs must navigate the tension between safeguarding ethical standards, such as avoiding the use of AI in sensitive areas like hiring, and leveraging AI to enhance efficiency and effectiveness \cite{asajile2024ai}.

A further technological challenge concerns \textit{trust in AI}. For instance, a study on AI in biodegradable package design in Bangladesh found that merely 35\% of consumers place trust in AI-generated solutions \cite{ulfy2025ai}. Similarly, research on AI as a decision-maker in public services suggests that public perceptions remain largely negative, despite some evidence that AI can increase perceptions of distributive justice and thereby strengthen donation intentions in specific groups \cite{yang2024aijustice}. As most NGOs rely heavily on donations, such attitudes can directly influence the willingness to adopt AI \cite{dube2024factors}. At the same time, NGOs face considerable pressure to maintain public trust and transparency, making them particularly vulnerable to reputational damage \cite{huang2024tech, hahn2025strategic}. Such damage not only undermines the organizations' credibility but can also have legal consequences and risk their future funding \cite{kravchuk2025ethical}. 

\subsubsection{Challenges in the Organization Context}

The organization context of the TOE-framework refers to the internal characteristics of an organization that influence its ability and readiness to adopt and implement new technologies. For NGOs, \textit{financial constraints} fall within this category. Most NGOs have limited budgets that primarily stem from donations, grants, or government funding, which are more volatile than commercial revenue streams \cite{huang2024tech}. Since the acquisition and integration of AI systems can be costly, financial limitations are often cited as a major barrier to adoption \cite{toplic2020ai, dutta2024machine, faruq2024ai, dube2024factors, efthymiou2023role, ghani2021ml, elamin2024ai, scutto2025empowerment}. NGOs must therefore weigh the potential benefits of AI against the upfront cost of its implementation \cite{krause2025ai}. This challenge is not limited to low or lower-middle income countries but also remains central in higher-income contexts like Slovenia or the US \cite{nejc2025role, unesco2023sal}. As a result, NGOs might be especially motivated to experiment with affordable AI-generated solutions \cite{arango2023aiads}. However, in contrast to the mainstream view, some scholars suggest that free versions of AI applications may already be sufficient for many NGOs, thereby partially contradicting the assumption that AI adoption is hindered due to financial constraints \cite{hahn2025strategic}.

These financial constraints within NGOs are further amplified by the challenge of \textit{competition for funding}. Contemporary research notes that organizations that fail to embrace emerging technologies like AI that increase efficiency risk falling behind their competition, thereby diminishing future funding opportunities \cite{asajile2024ai, dube2024factors}. Larger charities, with their highly skilled teams and greater financial resources, possess a substantial capitalistic advantage over smaller organizations \cite{hansen2025ai}. This dynamic risks creating a “winner-takes-all” environment in which well-resourced NGOs consolidate their dominance, potentially reinforcing structural inequalities and worsening disparities within the non-profit sector.

Another crucial organizational challenge is the \textit{lack of awareness of AI or an unwillingness for its adoption} \cite{toplic2020ai, dube2024factors, efthymiou2023role, atalay2025ai}. For instance, many NGOs in Tanzania perceive AI adoption as a potential threat to their human resource practices \cite{asajile2024ai}, while a survey of NGOs in Zambia found that 56.7\% of respondents had limited or no familiarity with the technology \cite{dube2024factors}. This unwillingness is often driven by employees fearing replacement by AI \cite{hahn2025strategic, popescu2024impact} and a lack in leadership and management support \cite{dube2024factors}. Moreover, NGOs depend extensively on the individual motivation of their staff \cite{dutta2024machine}, many of whom already juggle multiple and overlapping responsibilities, particularly in smaller organizations, further constraining the capacity to experiment with or integrate new technologies.\cite{dutta2024machine}. 

As AI implementation usually requires a minimum of technical familiarity with itself,  \textit{lacking expertise} can be another major organizational barrier to adoption \cite{dube2024factors}. Many NGOs do not have what they believe to be adequate expertise in AI and a lack of skilled workforce with contextual understanding \cite{ahatsi2025humanitarian, faruq2024ai, sandberg2025ai, weber2024emergency, nethope2023amplifying, dubey2022impact, ahatsi2025resilience}. Adopting a knowledge-based view Scutto et al. argue that value creation of humanitarian AI actions is still largely determined by individual skills rather than collective knowledge \cite{scutto2025empowerment}. As AI technologies continue to grow in complexity, this expertise gap is becoming an increasingly critical obstacle \cite{elamin2024ai}. Interestingly, this effect does not only appear in low or lower-middle income countries but also in high-income contexts like Slovenia \cite{nejc2025role}.

Another major issue in the organization context are \textit{organizational challenges}. Many NGOs report that they do not feel like they have clear organizational policies regarding AI \cite{sandberg2025ai}. Additionally, the multitude of solutions and approaches to choose from, with roadmaps and pricing mechanisms that are not clearly defined yet, make the selection of the “right” solution a challenge \cite{nethope2023amplifying}. Furthermore, NGOs are often forced to adopt ad hoc information management solutions to react to crises, which leads to a loss of historic data, the need to "reinvent the wheel" each time and difficulties with setting up an AI infrastructure \cite{nethope2024empowering}. While these challenges persist, researchers like Ahatsi et al. argue that they remain marginal compared to technical and financial constraints \cite{ahatsi2025resilience}.

Within the organizational context, the \textit{size of an NGO} significantly shapes the AI adoption journey. In contrast to moderate or small sized NGOs, large organizations usually have an advanced IT infrastructure, a culture supportive of technological innovation and stronger donor support for AI initatives \cite{dube2024factors}. These advantages enable them to specifically hire technical experts, and they possess a clearer understanding of what they want from someone in this position \cite{niranjana2023ai}. They also have the capacity to actively initiate cooperations with experts from industry and academia \cite{toplic2020ai}, something that due to their resource constraints, small and moderate size NGOs sometimes lack. Moreover, more traditional or smaller entities are at risk in an artificial intelligence economy, facing greater challenges in adapting to technological transformations and maintaining competitiveness \cite{popescu2024impact}. Cooperations between large NGOs like NetHope also have the capacity to develop resources like the Artificial Intelligence Ethics for Non-Profit Toolkit, the Humanitarian AI Code of Conduct or the Data Governece Toolkit: A guide to implementing data governance in nonprofits \cite{weber2024emergency} which supports compliance with ethical and legal standards.

On the other hand, small NGOs often juggle limited budgets and staff while attempting to achieve significant social impact \cite{krause2025ai}. Their adoption of AI is further constrained by comparatively weaker digital infrastructures \cite{dube2024factors}, as well as the need to plan for scalability and continuous adaptation despite resource shortages \cite{krause2025ai}.

\subsubsection{Challenges in the Environment Context}

Within the TOE framework, the environmental context refers to external factors that influence an organizations' technology adoption. For example, \textit{legal challenges} can play a crucial role, as NGOs have to adhere to complex local and global legislative guidelines regarding the usage of AI \cite{krause2025ai}. In many places, however, the absence or inadequacy of legal frameworks can itself hinder the widespread adoption of AI technologies \cite{atalay2025ai}.

A further environmental challenge is the \textit{lack of collaborations}, with academic, government and industry partners, which can hinder AI adoption in NGOs \cite{dube2024factors, faruq2024ai}. In general, structural weaknesses in public–private sector collaboration limit the societal benefits of technological transfer. Some countries, such as Morocco, are now actively seeking to encourage such partnerships \cite{atalay2025ai}. The collaboration between the charity sector and government bodies also often lags behind, as many NGOs encounter difficulties in accessing government data \cite{nethope2024empowering}. Moreover, the absence of a common standard for a systematic information management process among different NGOs, frequently leads to individual flawed adoption and ad hoc cooperations that are not persistent \cite{nethope2024empowering}.

Finally, \textit{geographic context} plays a crucial role in shaping AI adoption in NGOs. This relates back to the digital divide, which represents a structural gap in infrastructure, capabilities, and knowledge between different nations which can significantly hinder adoption \cite{weber2024emergency}. In some regions, limited internet penetration can constrain data handling \cite{nethope2024empowering}, while NGOs may need to charge for AI-based services due to high costs, reducing accessibility in low income contexts as a result \cite{unesco2023sal}. Furthermore, resource constraint locations often require specialized solutions, for instance through edge computing, that are not publicly or cheaply available \cite{nair2022adviser}. Additionally, discrepancies between policy objectives and their execution, especially in enforcement across various geographic areas \cite{ulfy2025ai} further worsen geographic disparities. 

On the other hand, high-income nations, like Qatar, have the capacity to make support for AI integration in NGOs a priority \cite{bahameish2023ai}. A study investigating generative AI use in NGO owned schools in Egypt also emphasized the importance of localization, including the use of local language, to enhance AI’s impact on local communities \cite{soudi2023generative}. However, biases in the available training data can structurally hinder localization efforts. For instance, Shrma et al. noted an insufficient number of datasets for non-English tools, showcasing how certain areas can be structurally disadvantaged in the usage of AI \cite{sharma2021ocr}. In addition, a dataset-driven study observed that ecological NGOs in areas such as Southern Europe, Northern and Middle Africa, Central Asia, Polynesia, and Micronesia have yet to adopt AI tools \cite{kazemeini2025dataset}. While these findings are based on limited evidence, the authors theorize that these regions face barriers including insufficient technological infrastructure, skill gaps, funding constraints, and restricted data access. 

\subsubsection{Proposed Solutions for AI Adoption}

To solve these numerous challenges, literature proposes a set of guidelines and solutions. A common starting point is capacity building, which includes improving data quality \cite{ahatsi2025humanitarian, ahatsi2025humanitarian,nethope2023amplifying} and investments in comprehensive staff training \cite{ulfy2025ai,dube2024factors,faruq2024ai,krause2025ai, ahatsi2025resilience, popescu2024impact}. This ensures that AI systems operate on reliable inputs and that staff develops the necessary skills and confidence to engage with new technologies. Additionally, targeted funding and resource allocations dedicated specifically to AI can provide the financial stability required to support these initiatives \cite{dube2024factors}.

Because of the high stake nature under which many NGOs operate, AI solutions must be tested, evaluated and refined before deployment \cite{weber2024emergency}. Such processes enable organizations to better understand both the potential and limitations of AI \cite{efthymiou2023role, dube2024factors}. A common strategy is to begin with small-scale pilot projects and incrementally build up on initial findings \cite{toplic2020ai}. Stakeholder involvement is essential in this phase, ensures that refinements are grounded in practical insights and organizational needs \cite{elamin2024ai}. For instance, NGOs in Tanzania utilized qualitative measures to understand how and when AI should be integrated in their hiring process \cite{asajile2024ai}. This incremental approach is paramount as it ensures both reliability and ethical use throughout all stages of the AI lifecycle \cite{krause2025ai, kravchuk2025ethical, alnamrouti2022ai, kbah2023understanding}. Furthermore, regular audits can support decision-making on whether existing solutions suffice or further development is necessary \cite{nethope2023amplifying}.

Another important insights is that open-source solutions are sometimes already sufficient for the intended use case of the NGO \cite{nejc2025role}. In a semi-strucutred interview with ten unspecified NGOs Popescu et al. found that most organizations currently primarily rely on open-source chatbots \cite{popescu2024impact}. The use of low-cost intelligent systems can not only help adoption \cite{faruq2024ai} but also keep the product at scale \cite{mate2022bandits}. However, choosing the right tool remains essential. For instance, Kraus developed an overview for NGOs regarding the pros and cons of available accounting solutions \cite{krause2025ai}. In addition, some AI providers offer discounted or free rates for social organizations, thereby helping to overcome financial boundaries \cite{faruq2024ai}. As the cost of AI is anticipated to decrease further, NGOs such as Khan Academy remain optimistic about the prospect of wider adoption \cite{unesco2023sal}.

With respect to AI-generated content, most studies advocate for a balanced approach. Research indicates that the use of AI-generated images can negatively impact donation intentions \cite{arango2023aiads}. However, making ethical motives salient or using these images under extraordinary circumstances which require timely response, was shown to lead to similar donation outcomes as using real images \cite{arango2023aiads}. Furthermore, deploying functional (e.g. ease of use) and emotional features (e.g. emotional connection) in interactive chatbots was shown to increase trust, which in turn correlates with stronger social media engagement and higher willingness to donate \cite{cheng2025aifeed}.

As practical guidance, research emphasizes the importance of ensuring data anonymity and security through a privacy-by-design approach \cite{kravchuk2025ethical, blasi2024eumigratool} and engagement with critical theory \cite{sandberg2025ai}. To preempt biases, some studies suggest removing sensitive attributes, such as gender-identifying features, when they are not strictly necessary \cite{avagyan2020utilizing}. At the same time, it is important to recognize that AI should not be viewed as an end in itself. In many cases non-AI solutions fully suffice for their intended use cases \cite{pai2023ngo}. Ultimately, what matters most is that applied solutions are genuinely useful and well scoped, while still allowing for scalability and adaptability \cite{elamin2024ai, nethope2023amplifying}.

There are also many calls for a unified, inclusive AI governance framework that prioritizes ethical considerations, ensures adequate resource allocation, strengthens education and training, and fosters collaboration between governments, NGOs, and academic institutions \cite{nejc2025role, ghani2021ml}. At the same time, research often appeals to policymakers to create a supportive environment for AI adoption in NGOs, by for instance investing in digital infrastructure \cite{asajile2024ai}, or creating concrete legals structures to guide the adoption process \cite{atalay2025ai,ahatsi2025resilience}.

Another crucial factor in addressing challenges is the importance of the environment and sociocultural context \cite{atalay2025ai, rahman2021development}. Pereira \& Shafique argue, that NGOs should focus on their unique resources and adapt their strategies to local conditions in order to effectively implement AI into their supply chains \cite{pereira2024ai}. Furthermore, individual geographic constraints have to be assessed by local NGO, as external collaborators often lack the contextual knowledge required \cite{nair2022adviser}. To enhance access and distributive justice, AI solutions should also be localized, both by providing interfaces in local languages and by ensuring remote accessibility in underdeveloped regions \cite{kravchuk2025ethical, soudi2023generative, ahatsi2025humanitarian}.

What emerges most strongly from the literature is the emphasis on partnerships and collaborations as a prerequisite for successful AI adoption in NGOs \cite{dube2024factors, faruq2024ai}. For one, this includes inter-NGO cooperation to develop shared  ethical framework for AI use \cite{ahatsi2025humanitarian} like the Artificial Intelligence Ethics for Nonprofits Toolkit \cite{weber2024emergency}. Additionally, collaboration with government bodies through data sharing, and joint initiatives improves the effectiveness of AI \cite{alex2023designing, nethope2024empowering}. However, when acting as data providers to government bodies, NGOs have to maintain neutrality \cite{iazzolino2025trading}. To profit from state-of-the-art AI solutions, a close collaboration with technology companies also remains beneficial \cite{ahatsi2025humanitarian, faruq2024ai, pai2023ngo, ahatsi2025resilience, kbah2023understanding}. Improving public–private sector collaborations is identified as one of the key solutions for AI Adoption in NGOs \cite{atalay2025ai}. Academic collaborations provide an additional pathway, helping NGOs experiment with and evaluate AI in practice \cite{blasi2024eumigratool, tingzon2020mapping, tabar2022forecasting, thalor2024greenmapper, mate2022bandits, alhindi2020vehicle}. This systematic literature review includes a wide spectrum of such collaborations, ranging from localized projects in low-income contexts \cite{grass2023ml} to long-term, large-scale partnerships such as ADVISER \cite{nair2022adviser}, illustrating the diversity of approaches available to NGOs.

\section{Discussion \& Implications}
This systematic literature review examined 65 documents to answer two central questions: How NGOs currently utilize AI (RQ1), and what challenges and solutions shape their adoption processes (RQ2). In the following we interpret our findings, emphasize theoretical and practical implications, state limitations, and discuss future research directions.

While we observed a growing research interest in NGOs adoption of AI, utilizing the DOI framework we placed the current adoption stage, depending on context, between early adoptor and early majority \cite{kazemeini2025dataset, dube2024factors}. Across the literature, NGOs utilize AI in the following six ways: 
\begin{itemize}
 \item \textit{Engagement}: AI tools for outreach and conversational support (e.g. interactive chatbots) that scale communication and personalize donor and beneficiary interactions.
 \item \textit{Optimization}: AI applied to logistics, resource allocation and fundraising efficiency, for example in vehicle routing and donor matching.
 \item \textit{Decision-support}: Predictive analytics and dashboards that inform prioritization and operational decisions in crises or program targeting.
 \item \textit{Management}: Automation for back-office functions like finance, procurement, HR, or volunteer coordination.
 \item \textit{Creative}: Generative AI for content production, educational material curation and campaign assets.
 \item \textit{Predictive}: Models usually using geospatial or socio-economic data to forecast displacement, disasters, or risk (early-warning systems).
\end{itemize}
Generally, large NGOs are more capable to develop in-house and complex projects \cite{dube2024factors, niranjana2023ai}, while smaller organizations tend to use off-the-shelf or open-source tools \cite{krause2025ai}. This resource and capability difference helps explain the uneven diffusion of adoption under the DOI model and underscores the risk of inequity within the NGO sector.

Utilizing the TOE-framework, we identified multiple challenges to AI adoption in NGOs. Within the technological context, challenges include limited data availability and diverse ethical challenges such as fairness, accountability, security, the need to keep humans in the loop, and the risk of overreliance on technology. Furthermore, concerns with trust and user acceptability further complicate adoption. In the organizational context, the most pressing obstacles are financial constraints, especially among smaller NGOs \cite{arango2023aiads}, which contribute to intensified competition for funding. Additionally, research indicates a lack of awareness, willingness, expertise and organizational preparedness. Size-specific limitations, such as restricted infrastructure and staff capacity, place smaller NGOs at a further disadvantage \cite{dube2024factors}. Finally, challenges in the environment context include inadequate or absent legal frameworks, structurally disadvantaged digital infrastructure, skill gaps, and restricted data accessibility in low- and lower-middle-income countries \cite{nair2022adviser}, alongside weak or fragmented partnerships.

The literature proposes several solution pathways. These include initial capacity building through improved data governance and staff training, pilot projects, incremental approaches and leveraging open-source solutions that require less upfront investments compared to in-house products. The suitability of these solutions is strongly shaped by organizational size and geographic context. However, across studies, the most consistently emphasized recommendation is the need for stronger collaborations between NGOs and external actors such as governments, industry, and academia \cite{faruq2024ai}.

Based on available literature, NGOs should approach AI adoption through a stepwise process that includes initial need assessment, running small pilot projects, ensuring secure data governance, and scaling solutions in collaboration with partners. Donors are advised to fund capacity building, underwrite shared infrastructure, and support the localization of the technology. Policymakers can accelerate adoption by investing in digital infrastructure, create legal clarity on data use, and incentive public-private partnerships. Furthermore, industry partners can contribute by offering discounted or free access for NGOs, developing tailored AI applications, and providing privacy-preserving hosting options.

We acknowledge that this study faces several limitations. First, by excluding non-English literature, we might introduce publication and language bias. Second, although both authors worked independently, regularly cross-checked their results, and followed a predefined research protocol, the small author team still leaves room for potential bias. Third, the sample of investigated studies is skewed toward lower-middle and high-income countries as well as larger NGOs, with many studies positioning NGOs merely as end users or data providers rather than active collaborators. Fourth, the heterogeneity in study designs prevented a meta-analysis, which means we defaulted to narrative and thematic synthesis, which includes inherently subjective elements. Finally, we note that while we also searched for grey literature, there might be possible misses or emerging projects not yet published. As a result, this paper primarily views the issue from an academic perspective. 

Future research should prioritize empirical studies of AI adoption in small NGOs and low-income countries to fill in existing research gaps. Additionally, longitudinal studies that track AI adoption over time to test the DOI trajectories and comparative evaluations of open-source vs paid tools in NGOs seem to be desirable. Finally, further investigations into localization (languages, cultural adaptation) of AI tools would provide a foundation for more effective and context-sensitive implementation.

\section{Conclusion}
This review examined how NGOs adopt and use AI, with a focus on utilization patterns, barriers, and solutions across different organizational and geographic contexts. In sum, it shows that AI is currently being used by NGOs primarily for engagement, optimization, decision-support, management, creative, and predictive purposes, with structured uptake concentrated among larger organizations (RQ1). The main barriers are technical capacity, organizational resources, and environmental constraints such as funding and infrastructure, while feasible solutions include partnerships, capacity-building, and governance frameworks that ensure ethical and context-sensitive use (RQ2). Overall, AI is promising, but adoption remains uneven. Over the next five years, policies should prioritize equitable capacity building, supportive infrastructure, and ethical guidelines to ensure smaller NGOs and those from lower-income contexts are not left behind. At the same time, research should expand to underrepresented contexts and investigate longitudinal patterns of AI adoption to inform sustainable strategies.

\section{Statements and Declarations}
The authors have no competing interests to declare that are relevant to the content of this article. Additionally, they did not receive support from any organization for the submitted work.

\section{Acknowledgements}
Special thanks is attributed to Grow Ghana, a Ghanian NGO, that was one of the main motivations to conduct this work.

In some parts, we used ChatGPT version 5 for grammar polishing and style fixes. These contributions were minor and all content was reviewed and edited by the authors.


\printbibliography

@online{metreau2024world,
  author    = {Eric Metreau and Kathryn Elizabeth Young and Shwetha Grace Eapen},
  title     = {World Bank country classifications by income level for 2024–2025},
  year      = {2024},
  organization = {World Bank Group},
  url       = {https://blogs.worldbank.org/en/opendata/world-bank-country-classifications-by-income-level-for-2024-2025},
  note      = {Accessed: 2025-09-02}
}

@article{cooke2012beyond,
  author       = {Cooke, Alison and Smith, David and Booth, Andrew},
  title        = {Beyond PICO: The SPIDER Tool for Qualitative Evidence Synthesis},
  journal      = {Qualitative Health Research},
  year         = {2012},
  volume       = {22},
  number       = {10},
  pages        = {1435--1443},
  doi          = {10.1177/1049732312452938}
}

@article{richardson1995well,
  title={The well-built clinical question: a key to evidence-based decisions.},
  author={Richardson, W Scott and Wilson, Mark C and Nishikawa, Jim and Hayward, Robert S},
  journal={ACP journal club},
  volume={123},
  number={3},
  pages={A12--3},
  year={1995}
}

@article{martin2021google,
  title={Google Scholar, Microsoft Academic, Scopus, Dimensions, Web of Science, and OpenCitations’ COCI: a multidisciplinary comparison of coverage via citations},
  author={Mart{\'\i}n-Mart{\'\i}n, Alberto and Thelwall, Mike and Orduna-Malea, Enrique and Delgado L{\'o}pez-C{\'o}zar, Emilio},
  journal={Scientometrics},
  volume={126},
  number={1},
  pages={871--906},
  year={2021},
  publisher={Springer},
  doi = {10.1007/s11192-020-03690-4}
}

@article{adams2017shades,
  title={Shades of grey: guidelines for working with the grey literature in systematic reviews for management and organizational studies},
  author={Adams, Richard J and Smart, Palie and Huff, Anne Sigismund},
  journal={International journal of management reviews},
  volume={19},
  number={4},
  pages={432--454},
  year={2017},
  publisher={Wiley Online Library},
  doi = {10.1111/ijmr.12102}
}

@article{hirt2023citation,
  title={Citation tracking for systematic literature searching: A scoping review},
  author={Hirt, Julian and Nordhausen, Thomas and Appenzeller-Herzog, Christian and Ewald, Hannah},
  journal={Research Synthesis Methods},
  volume={14},
  number={3},
  pages={563--579},
  year={2023},
  publisher={Wiley Online Library},
  doi = {10.1002/jrsm}
}

@article{pigott2020methodological,
  title={Methodological guidance paper: High-quality meta-analysis in a systematic review},
  author={Pigott, Terri D and Polanin, Joshua R},
  journal={Review of Educational Research},
  volume={90},
  number={1},
  pages={24--46},
  year={2020},
  publisher={SAGE Publications Sage CA: Los Angeles, CA},
  doi = {10.3102/0034654319877153}
}

@article{walters2025comparing,
  title={Comparing conventional and alternative mechanisms of discovering and accessing the scientific literature},
  author={Walters, William H},
  journal={Proceedings of the National Academy of Sciences},
  volume={122},
  number={27},
  pages={e2503051122},
  year={2025},
  publisher={National Academy of Sciences},
  doi = {10.1073/pnas.2503051122}
}

@article{shams2025ai,
  title={AI and the quest for diversity and inclusion: a systematic literature review},
  author={Shams, Rifat Ara and Zowghi, Didar and Bano, Muneera},
  journal={AI and Ethics},
  volume={5},
  number={1},
  pages={411--438},
  year={2025},
  publisher={Springer},
  doi = {10.1007/s43681-023-00362-w}
}

@article{page2021prisma,
  title={The PRISMA 2020 statement: an updated guideline for reporting systematic reviews},
  author={Page, Matthew J and McKenzie, Joanne E and Bossuyt, Patrick M and Boutron, Isabelle and Hoffmann, Tammy C and Mulrow, Cynthia D and Shamseer, Larissa and Tetzlaff, Jennifer M and Akl, Elie A and Brennan, Sue E and others},
  journal={bmj},
  volume={372},
  year={2021},
  publisher={British Medical Journal Publishing Group},
  doi = {10.1136/bmj.n71}
}

@misc{CASP2018,
  author       = {{Critical Appraisal Skills Programme (CASP)}},
  title        = {CASP Qualitative Studies Checklist},
  year         = 2018,
  url          = {https://casp-uk.net/casp-tools-checklists/qualitative-studies-checklist/},
  note         = {Accessed: 2025-08-16}
}

@article{mohammed2016meta,
  title={Meta-synthesis of qualitative research: the challenges and opportunities},
  author={Mohammed, Mohammed A and Moles, Rebekah J and Chen, Timothy F},
  journal={International journal of clinical pharmacy},
  volume={38},
  number={3},
  pages={695--704},
  year={2016},
  publisher={Springer},
  doi = {10.1007/s11096-016-0289-2}
}

@techreport{AI_Market_2025_2030,
  title        = {Artificial Intelligence Market Size, Share \& Trends Analysis Report By Solution, By Technology (Deep Learning, Machine Learning, NLP, Machine Vision, Generative AI), By Function, By End-Use, By Region, And Segment Forecasts, 2025--2030},
  author       = {{Grand View Research, Inc.}},
  institution  = {Grand View Research},
  year         = {2025},
  type         = {Industry report},
  note         = {Forecast period: 2025--2030},
  url          = {https://www.grandviewresearch.com/industry-analysis/artificial-intelligence-ai-market},
  urldate      = {2025-08-21}
}

@article{madimi2023,
title = {Integration of AI-Powered Analytics for Donor Behavior Insights},
author = {Satwik Mamidi},
year = {2023},
journal = {Progress in Medical Sciences},
volume = {7},
number = {2},
pages = {1--4},
doi = {10.47363/PMS/2023(7)E127}
}

@article{godefroid2024identifying,
  title={Identifying key barriers to nonprofit organizations' adoption of technology innovations},
  author={Godefroid, Marie-E and Plattfaut, Ralf and Niehaves, Bj{\"o}rn},
  journal={Nonprofit Management and Leadership},
  volume={35},
  number={1},
  pages={237--259},
  year={2024},
  publisher={Wiley Online Library},
  doi = {10.1002/nml.21609}
}

@article{eng2024impact,
  title={The Impact of AI-Driven Performance Evaluation on Organizational Outcomes in Kenya: A Systematic Literature Review},
  author={Eng’airo, Pamela},
  journal={Journal of Information and Technology},
  volume={8},
  number={2},
  pages={1--15},
  year={2024},
  doi = {10.53819/81018102t7040 }
}

@article{schiff2021ai,
  title={AI ethics in the public, private, and NGO sectors: A review of a global document collection},
  author={Schiff, Daniel and Borenstein, Jason and Biddle, Justin and Laas, Kelly},
  journal={IEEE Transactions on Technology and Society},
  volume={2},
  number={1},
  pages={31--42},
  year={2021},
  publisher={IEEE},
  doi = {10.1109/TTS.2021.3052127}
}

@article{tomazevic2024organizational,
  title={Organizational Enablers of Artificial Intelligence Adoption in Public Institutions: A Systematic Literature Review},
  author={Tomazevic, Nina and Murko, Eva and Aristovnik, Aleksander},
  journal={Cent. Eur. Pub. Admin. Rev.},
  volume={22},
  pages={109},
  year={2024},
  publisher={HeinOnline},
  doi = {10.17573/cepar.2024.1.05}
}

@article{taboada2023artificial,
  title={Artificial intelligence enabled project management: a systematic literature review},
  author={Taboada, Ianire and Daneshpajouh, Abouzar and Toledo, Nerea and De Vass, Tharaka},
  journal={Applied Sciences},
  volume={13},
  number={8},
  pages={5014},
  year={2023},
  publisher={MDPI},
  doi = {10.3390/app13085014}
}

@article{marquez2016relevance,
  title={The relevance of organizational structure to NGOs’ approaches to the policy process},
  author={Marquez, Luz M Mu{\~n}oz},
  journal={VOLUNTAS: International Journal of Voluntary and Nonprofit Organizations},
  volume={27},
  number={1},
  pages={465--486},
  year={2016},
  publisher={Springer},
  doi = {10.1007/s11266-015-9555-5}
}

@book{russell2020aima,
  title     = {Artificial Intelligence: A Modern Approach},
  author    = {Russell, Stuart J. and Norvig, Peter},
  year      = {2020},
  edition   = {4},
  publisher = {Pearson},
  address   = {Upper Saddle River, NJ}
}

@book{tornatzky1990processes,
  title     = {The Processes of Technological Innovation},
  author    = {Tornatzky, Louis G. and Fleischer, Mitchell},
  year      = {1990},
  publisher = {Lexington Books},
  address   = {Lexington, MA}
}

@incollection{baker2011technology,
  author    = {Jeff Baker},
  title     = {The Technology–Organization–Environment Framework},
  booktitle = {Information Systems Theory: Explaining and Predicting Our Digital Society},
  editor    = {Y. K. Dwivedi and M. R. Wade and S. L. Schneberger},
  publisher = {Springer},
  year      = {2011},
  volume    = {28},
  pages     = {231--245},
  doi       = {10.1007/978-1-4419-6108-2_12},
}

@article{zhu2005post,
  title={Post-adoption variations in usage and value of e-business by organizations: cross-country evidence from the retail industry},
  author={Zhu, Kevin and Kraemer, Kenneth L},
  journal={Information systems research},
  volume={16},
  number={1},
  pages={61--84},
  year={2005},
  publisher={INFORMS},
  doi = {10.1287/isre.1050.0045}
}

@book{rogers2003diffusion,
  title     = {Diffusion of Innovations},
  author    = {Rogers, Everett M.},
  edition   = {5th},
  year      = {2003},
  publisher = {Free Press},
  address   = {New York}
}

@article{dubey2022impact,
  title = {Impact of artificial intelligence-driven big data analytics culture on agility and resilience in humanitarian supply chain: A practice-based view},
  journal = {International Journal of Production Economics},
  volume = {250},
  author = {Rameshwar Dubey and David J. Bryde and Yogesh K. Dwivedi and Gary Graham and Cyril Foropon},
  pages = {108618},
  year = {2022},
  note = {Special Issue celebrating Volume 250 of the International Journal of Production Economics},
  issn = {0925-5273},
  doi = {https://doi.org/10.1016/j.ijpe.2022.108618}
}

@inproceedings{castro2022ai,
author = {Castro, Francisco Enrique Vicente and DesPortes, Kayla and Payne, William and Bergner, Yoav and McDermott, Kathleen},
title = {AI + Dance: Co-Designing Culturally Sustaining Curricular Resources for AI and Ethics Education Through Artistic Computing},
year = {2022},
isbn = {9781450391955},
publisher = {Association for Computing Machinery},
address = {New York, NY, USA},
doi = {10.1145/3501709.3544275},
booktitle = {Proceedings of the 2022 ACM Conference on International Computing Education Research - Volume 2},
pages = {26–27},
numpages = {2},
location = {Lugano and Virtual Event, Switzerland},
series = {ICER '22}
}

@article{efthymiou2023role,
author = {Efthymiou, Iris-Panagiota and Alevizos, Antonios and Sidiropoulos, Symeon},
year = {2023},
month = {08},
pages = {1-7},
title = {The Role of Artificial Intelligence in Revolutionizing NGOs' Work},
volume = {2},
journal = {Journal of Politics and Ethics in New Technologies and AI},
doi = {10.12681/jpentai.35137}
}

@article{rathore2022sustainable,
  title        = {A Sustainable Model for Emergency Medical Services in Developing Countries: A Novel Approach Using Partial Outsourcing and Machine Learning},
  author       = {Rathore, N. and Jain, P. K. and Parida, M.},
  journal      = {Risk Management and Healthcare Policy},
  volume       = {15},
  pages        = {193--218},
  year         = {2022},
  doi          = {10.2147/RMHP.S338186},
  url          = {https://doi.org/10.2147/RMHP.S338186}
}

@article{grass2023ml,
  title        = {A Machine Learning Approach to Deal with Ambiguity in the Humanitarian Decision-Making},
  author       = {Grass, Emilia and Ortmann, Janosch and Balcik, Burcu and Rei, Walter},
  journal      = {Production and Operations Management},
  volume       = {32},
  number       = {9},
  pages        = {2956-2974},  
  year         = {2023},
  doi          = {10.1111/poms.14018},
  url          = {https://doi.org/10.1111/poms.14018},
  note         = {Handling Editor: Sushil Gupta}
}

@INPROCEEDINGS{tingzon2020mapping,
  author={Tingzon, Isabelle and Dejito, Niccolo and Flores, Ren Avell and De Guzman, Rodolfo and Carvajal, Liliana and Erazo, Katerine Zapata and Contreras Cala, Ivan Enrique and Villaveces, Jeffrey and Rubio, Daniela and Ghani, Rayid},
  booktitle={2020 IEEE / ITU International Conference on Artificial Intelligence for Good (AI4G)}, 
  title={Mapping New Informal Settlements Using Machine Learning and Time Series Satellite Images: An Application in the Venezuelan Migration Crisis}, 
  year={2020},
  pages={198-203},
  doi={10.1109/AI4G50087.2020.9311041}
}

@inproceedings{bahameish2023ai,
author = {Bahameish, Bedoor and Yaqot, Mohammed and Franzoi, R. and Menezes, Brenno},
year = {2023},
month = {04},
title = {Artificial Intelligence in Procurement: An Overview and Case Study of Qatar Foundation},
doi = {10.46254/EU05.20220146}
}

@article{pereira2024ai,
  title        = {The Role of Artificial Intelligence in Supply Chain Agility: A Perspective of Humanitarian Supply Chain},
  author       = {Pereira, Elisabeth T. and Shafique, Muhammad Noman},
  journal      = {Engineering Economics},
  volume       = {35},
  number       = {1},
  year         = {2024},
  doi          = {10.5755/j01.ee.35.1.32928},
}

@article{huang2024tech,
  title        = {Technology-Driven Financial Risk Management: Exploring the Benefits of Machine Learning for Non-Profit Organizations},
  author       = {Huang, H.},
  journal      = {Systems},
  volume       = {12},
  number       = {10},
  pages        = {416},
  year         = {2024},
  doi          = {10.3390/systems12100416},
  url          = {https://doi.org/10.3390/systems12100416}
}

@inproceedings{tabar2022forecasting,
author = {Tabar, Maryam and Jung, Wooyong and Yadav, Amulya and Chavez, Owen and Flores, Ashley and Lee, Dongwon},
year = {2022},
month = {07},
pages = {5144-5150},
title = {Forecasting the Number of Tenants At-Risk of Formal Eviction: A Machine Learning Approach to Inform Public Policy},
doi = {10.24963/ijcai.2022/715}
}

@inproceedings{nair2022adviser,
  title        = {{ADVISER}: AI-Driven Vaccination Intervention Optimiser for Increasing Vaccine Uptake in Nigeria},
  author       = {Nair, Vineet and Prakash, Kritika and Wilbur, Michael and Taneja, Aparna and Namblard, Corinne and Adeyemo, Oyindamola and Dubey, Abhishek and Adereni, Abiodun and Tambe, Milind and Mukhopadhyay, Ayan},
  booktitle    = {Proceedings of the International Joint Conference on Artificial Intelligence (IJCAI-22), AI for Good Track},
  year         = {2022},
  doi          = {10.48550/arXiv.2204.13663},
  url          = {https://doi.org/10.48550/arXiv.2204.13663}
}

@inproceedings{soudi2023generative,
author = {Soudi, Marwa and Ali, Esraa and Bali, Maha and Mabrouk, Nihal},
title = {Generative AI-Based Tutoring System for Upper Egypt Community Schools},
year = {2023},
isbn = {9798400716461},
publisher = {Association for Computing Machinery},
address = {New York, NY, USA},
url = {https://doi.org/10.1145/3633083.3633085},
doi = {10.1145/3633083.3633085},
booktitle = {Proceedings of the 2023 Conference on Human Centered Artificial Intelligence: Education and Practice},
pages = {16–21},
numpages = {6},
location = {Dublin, Ireland},
series = {HCAIep '23}
}

@article{niranjana2023ai,
  title        = {AI for Sustainable Development: Assessing Student Interest, Education, and Career Pathways},
  author       = {Niranjana, S.},
  journal      = {EPRA International Journal of Research and Development (IJRD)},
  volume       = {8},
  number       = {10},
  year         = {2023},
  issn         = {2455-7838},
  doi          = {10.36713/epra14795},
}

@article{nejc2025role,
author = {Brezovar, Nejc},
year = {2025},
month = {06},
pages = {11-30},
title = {The Role of Artificial Intelligence in NGOs: Challenges and Opportunities for Slovenia’s Information Society},
volume = {18},
journal = {NISPAcee Journal of Public Administration and Policy},
doi = {10.2478/nispa-2025-0002}
}

@article{blasi2024eumigratool,
  title        = {Developing AI Predictive Migration Tools to Enhance Humanitarian Support: The Case of EUMigraTool},
  author       = {Blasi Casagran, Cristina and Stavropoulos, Georgios},
  journal      = {Data \& Policy},
  volume       = {6},
  pages        = {e64},
  year         = {2024},
  doi          = {10.1017/dap.2024.76},
}

@article{amalraj2025ml,
  title        = {A machine learning approach to categorizing countries by socio-economic and health development factors using PCA, K-means, and silhouette scoring},
  author       = {Amalraj, Victoire and Vasuki, M. and Anitha, S.},
  journal      = {International Scientific Journal of Engineering and Management},
  volume       = {04},
  number       = {06},
  year         = {2025},
  month        = {June},
  doi          = {10.55041/ISJEM03977}
}

@article{atalay2025ai,
  title        = {Artificial Intelligence Technologies as Smart Solutions for Sustainable Protected Areas Management},
  author       = {Atalay, A. and Perkumienė, D. and Safaa, L. and Škėma, M. and Aleinikovas, M.},
  journal      = {Sustainability},
  volume       = {17},
  number       = {11},
  pages        = {5006},
  year         = {2025},
  doi          = {10.3390/su17115006}
}

@article{ahatsi2025humanitarian,
  title        = {Enhancing Humanitarian Supply Chain Resilience: Evaluating Artificial Intelligence and Big Data Analytics in Two Nations},
  author       = {Ahatsi, E. and Olanrewaju, O. A.},
  journal      = {Logistics},
  volume       = {9},
  number       = {2},
  pages        = {64},
  year         = {2025},
  doi          = {10.3390/logistics9020064}
}

@misc{ulfy2025ai,
  title        = {Integration of Artificial Intelligence in Biodegradable Plastic Packaging Design: Exploring Stakeholder Attitudes},
  author       = {Ulfy, Mohammad Arije and Haque, Ahasanul and Huda, Md. Nazmul},
  year         = {2025},
  month        = {May 20},
  howpublished = {SSRN Electronic Journal},
  doi          = {10.2139/ssrn.5367646},
}

@article{alex2023designing,
  title={Designing Artificial Intelligence Equipped Social Decentralized Autonomous Organizations for Tackling Sextortion Cases Version 0.7},
  author={Alex, Norta and Sotiris, Makrygiannis},
  journal={arXiv preprint arXiv:2312.14090},
  year={2023}
}

@article{asajile2024ai,
  title        = {Influence of Artificial Intelligence on Selection Stage of Recruitment in Tanzania: A Case of Selected NGOs in Kinondoni Municipality},
  author       = {Asajile, Michael and Komba, Silverius and Mbogo, Crispin},
  journal      = {International Journal of Research and Innovation in Social Science (IJRISS)},
  volume       = {8},
  number       = {12},
  pages        = {3406--3416},
  year         = {2024}
}

@article{thalor2024greenmapper,
  title        = {Green Mapper: An AI-Driven Initiative for Aerial Tree Mapping, Maintaining Environmental Balance},
  author       = {Thalor, M. and Khan, S. and Bhongale, S. and Bhosale, P. and Giri, A. and Shewale, S.},
  journal      = {Current Agricultural Research Journal},
  volume       = {12},
  number       = {2},
  year         = {2024},
  doi          = {10.12944/CARJ.12.2.34}
}

@misc{rafner2023crea,
  title        = {Crea.visions: A Platform for Casual Co-Creation with a Purpose Envisioning the Future through Human-AI Collaboration with Multiple Stakeholders},
  author       = {Rafner, J. F. and Zana, B. and Beolet, T. and Büyükgüzel, S. and Michel, E. and Maiden, N. and Risi, S. and Sherson, J.},
  year         = {2023},
  howpublished = {Open Access Report},
  note         = {\url{https://openaccess.city.ac.uk/id/eprint/30700/}}
}

@article{ghani2021ml,
  title        = {Machine Learning for Social Good: Applications in Non-Profit and Public Sectors},
  author       = {Ghani, Rayid},
  journal      = {American Journal of Machine Learning},
  volume       = {2},
  number       = {4},
  year         = {2021}
}

@article{kravchuk2025ethical,
  title        = {Ethical Implications of AI Applications in Nonprofit and Charity Sectors},
  author       = {Kravchuk, Y.},
  journal      = {COMPUTER-INTEGRATED TECHNOLOGIES: EDUCATION, SCIENCE, PRODUCTION},
  volume       = {58},
  pages        = {46--52},
  year         = {2025},
  doi          = {10.36910/6775-2524-0560-2025-58-06}
}

@inproceedings{dube2024factors,
author = {Dube, Macdonald and Dube, Sibusisiwe and Mutunhu Ndlovu, Belinda},
year = {2024},
month = {07},
title = {Factors Influencing the Adoption of AI Chatbots By Non- Governmental Organizations},
doi = {10.46254/EU07.20240155}
}

@inproceedings{faruq2024ai,
author = {Faruq, Omar and Haque, Shariful and Sufian, Mohammad and Hossain, Mir Abrar and Talukder, Tughlok and Shayed, Uddin},
year = {2024},
month = {09},
title = {AI-Driven Strategies for Enhancing Non-Profit Organizational Impact},
doi = {10.62127/aijmr.2024.v02i05.1088}
}

@inproceedings{sharma2021ocr,
  author={Sharma, Ruchi and Dave, Parv and Chaudhary, Jay},
  booktitle={2021 Fifth International Conference on I-SMAC (IoT in Social, Mobile, Analytics and Cloud) (I-SMAC)}, 
  title={OCR for Data Retrieval :An analysis and Machine Learning Application model for NGO social volunteering}, 
  year={2021},
  pages={422-427},
  doi={10.1109/I-SMAC52330.2021.9640890}
}

@article{alnamrouti2022ai,
  title        = {Do Strategic Human Resources and Artificial Intelligence Help to Make Organisations More Sustainable? Evidence from Non-Governmental Organisations},
  author       = {Alnamrouti, A. and Rjoub, H. and Ozgit, H.},
  journal      = {Sustainability},
  volume       = {14},
  number       = {12},
  pages        = {7327},
  year         = {2022},
  doi          = {10.3390/su14127327}
}

@inproceedings{kazemeini2025dataset,
author = {Kazemeini, Amirmohammad and Qi, Sipu and Ildem, Ece Ozen and Nassiri, Fargol and Namazi, Reza and Atici, Ceren Zeytinoglu and Kocak, Sedef Akinli},
title = {A Dataset-Driven Study of AI Opportunities in the Climate NGO Ecosystem},
year = {2025},
isbn = {9798400714849},
publisher = {Association for Computing Machinery},
address = {New York, NY, USA},
doi = {10.1145/3715335.3744894},
booktitle = {Proceedings of the 2025 ACM SIGCAS/SIGCHI Conference on Computing and Sustainable Societies},
pages = {780–785},
numpages = {6},
series = {COMPASS '25}
}

@article{kapuge2024optimizing,
  title={Optimizing AI recommendation algorithms for efficient matching of most needed beneficiaries with donors in Sri Lankan charity sector},
  author={Kapuge, Sethika Manumitha and Ginige, Thepul NDS},
  journal={DATA SCIENCE AND AI-I 1-6},
  pages={63},
  year={2024}
}

@misc{krause2025ai,
  title        = {AI Agents and Automation in Small Non-Profit Organizations' Accounting Functions},
  author       = {Krause, David},
  year         = {2025},
  month        = {January 04},
  howpublished = {SSRN Electronic Journal},
  doi          = {10.2139/ssrn.5082437}
}

@article{sammer2024aifeed,
  title        = {AI-FEED: Prototyping an AI-Powered Platform for the Food Charity Ecosystem},
  author       = {Sammer, Marcus and Seong, Kijin and Olvera, Norma and Gronseth, Susie L. and Anderson-Fletcher, Elizabeth and Jiao, Junfeng and Reese, Alison and Kakadiaris, Ioannis A.},
  journal      = {International Journal of Computational Intelligence Systems},
  volume       = {17},
  pages        = {259},
  year         = {2024},
  doi          = {10.1007/s44196-024-00656-9}
}

@article{kannan2025ai,
author = {Kannan, M.K.Jayanthi and Shukla, Shivansh and Parmar, Anshuman and AbhinavJha, and Singh, Aryan and Utkarsh, and B, Harish},
year = {2025},
month = {04},
pages = {6600-6613},
title = {AI-Driven Digital Volunteering @ Freelancing Website for NGO Collaboration with Online Talent Worldwide},
volume = {8},
doi = {10.15680/IJMRSET.2025.0804405}
}

@article{laylo2023impact,
  title={The impact of AI and information technologies on Islamic charity (zakat): Modern solutions for efficient distribution},
  author={Laylo, Karshiboyeva},
  journal={International Journal of Law and Policy},
  volume={1},
  number={5},
  pages={1--8},
  year={2023}
}

@article{defnizal2020ai,
  title        = {The Implementation of Artificial Intelligence in Charity Box at Mosque and Musholla as RFID Based Security System},
  author       = {Defnizal, D. and Ernes, R. N.},
  journal      = {Sinkron: Jurnal Dan Penelitian Teknik Informatika},
  volume       = {5},
  number       = {1},
  pages        = {35--42},
  year         = {2020},
  doi          = {10.33395/sinkron.v5i1.10594}
}

@inproceedings{hansen2025ai,
  title        = {Exclusion or Efficiency: Understanding Perspectives about AI Ethics Among Charity Workers in the United Kingdom},
  author       = {Hansen, Sakina},
  booktitle    = {Proceedings of the Fourth European Workshop on Algorithmic Fairness (EWAF'25)},
  series       = {Proceedings of Machine Learning Research},
  year         = {2025}
}

@misc{sievert2024ai,
  title        = {AI, Diversity and Trust in Digital Non-Profit Communication: Results of a Survey within the Major Traditional Churches of a Western European Country},
  author       = {Sievert, Holger and Inderhees, Marco},
  year         = {2024},
  month        = {March 07},
  howpublished = {SSRN Electronic Journal},
  doi          = {10.2139/ssrn.5087283}
}

@article{dutta2024machine,
  title        = {The machine/human agentic impact on practices in learning and development: A study across MSME, NGO and MNC organizations},
  author       = {Dutta, Debolina and Kannan Poyil, Anasha},
  journal      = {Personnel Review},
  volume       = {53},
  number       = {3},
  pages        = {791--815},
  year         = {2024},
  doi          = {10.1108/PR-09-2022-0658}
}

@article{sandberg2025ai,
  title        = {Addressing the promise and peril of AI for nonprofit management through a data feminist pedagogy},
  author       = {Sandberg, B. and Wasif, R. and Hand, L. C.},
  journal      = {Journal of Public Affairs Education},
  volume       = {31},
  number       = {2},
  pages        = {192--212},
  year         = {2025},
  doi          = {10.1080/15236803.2025.2475589}
}

@article{cheng2025aifeed,
  title        = {Leveraging Artificial Intelligence–Powered Chatbots for Nonprofit Organizations: Examining the Antecedents and Outcomes of Chatbot Trust and Social Media Engagement},
  author       = {Cheng, Yang and Wang, Yuan},
  journal      = {Journal of Philanthropy},
  volume       = {30},
  number       = {1},
  pages        = {Article e70013},
  year         = {2025},
  doi          = {10.1002/nvsm.70013}
}

@incollection{ hahn2025strategic,
 title = {From Data to Donors: Can AI Reshape Fundraising Strategies},
 author = {Hahn, Luisa Regina and Hartmann, Lina and Knjasew, Anita and Weißer, Felicia Maria},
 editor = {Godulla, Alexander and Decker, Fabienne and Hartmann, Lina and Pankoke, Cinja and Pecher, Meriel and Pütter, Sarah and Santangelo, Chiara},
 year = {2025},
 booktitle = {Strategic Communication in Disruptive Times: How Sociopolitical Polarization, Virtual Media and AI Reshape Organizational Communication},
 pages = {115-137},
 address = {Leipzig},
 urn = {https://nbn-resolving.org/urn:nbn:de:0168-ssoar-102802-8}
}

@inproceedings{mate2022bandits,
  title        = {Field Study in Deploying Restless Multi-Armed Bandits: Assisting Non-profits in Improving Maternal and Child Health},
  author       = {Mate, A. and Madaan, L. and Taneja, A. and Madhiwalla, N. and Verma, S. and Singh, G. and Hegde, A. and Varakantham, P. and Tambe, M.},
  booktitle    = {Proceedings of the AAAI Conference on Artificial Intelligence},
  volume       = {36},
  number       = {11},
  pages        = {12017--12025},
  year         = {2022},
  doi          = {10.1609/aaai.v36i11.21460}
}

@article{joshi2025smart,
author = {Joshi, Divija},
year = {2025},
month = {05},
pages = {1-9},
title = {SmartNGO: An Integrated Platform for Managing Volunteers and Events},
volume = {09},
journal = {INTERNATIONAL JOURNAL OF SCIENTIFIC RESEARCH IN ENGINEERING AND MANAGEMENT},
doi = {10.55041/IJSREM46660}
}

@INPROCEEDINGS{pai2023ngo,
  author={Pai, Akanksha A and Kumar P, Ramakanth and Thomas, Sharon and D, Pratiba},
  booktitle={2023 7th International Conference on Computation System and Information Technology for Sustainable Solutions (CSITSS)}, 
  title={NGO CONNECT: Technology for Non-Profit Organisation Management}, 
  year={2023},
  pages={1-6},
  doi={10.1109/CSITSS60515.2023.10334076}
}

@article{elamin2024ai,
  title        = {Modernizing the Charitable Sector through Artificial Intelligence: Enhancing Efficiency and Impact},
  author       = {Elamin, Mustafa Osman I.},
  journal      = {Journal of Ecohumanism},
  volume       = {3},
  number       = {4},
  pages        = {3426--3443},
  year         = {2024},
  publisher    = {Transnational Press London}
}

@article{yang2024aijustice,
  title        = {Understanding the impact of artificial intelligence on the justice of charitable giving: The moderating role of trust and regulatory orientation},
  author       = {Yang, Chen and Yang, Yi and Zhang, Yuezi},
  journal      = {Journal of Consumer Behaviour},
  volume       = {23},
  number       = {5},
  pages        = {2624--2636},
  year         = {2024},
  month        = {September},
  doi          = {10.1002/cb.2365}
}

@article{arango2023aiads,
  title        = {Consumer responses to AI-generated charitable giving ads},
  author       = {Arango, L. and Singaraju, S. P. and Niininen, O.},
  journal      = {Journal of Advertising},
  volume       = {52},
  number       = {4},
  pages        = {486--503},
  year         = {2023},
  doi          = {10.1080/00913367.2023.2183285}
}

@article{alhindi2020vehicle,
author = {Alhindi, Ahmad and Alsaidi, Abrar and Alasmary, Waleed and Alsabaan, Maazen},
year = {2020},
month = {03},
pages = {680},
title = {Vehicle Routing Optimization for Surplus Food in Nonprofit Organizations},
volume = {11},
journal = {International Journal of Advanced Computer Science and Applications},
doi = {10.14569/IJACSA.2020.0110384}
}

@article{unesco2023sal,
  author  = {Khan, Sal and Savolainen, Anuliina},
  title     = {Sal Khan: "I see AI as an additional tool, but a very powerful one"},
  journal   = {The UNESCO Courier},
  year      = {2023},
  volume    = {4},
  pages     = {12--14},
  url       = {https://unesdoc.unesco.org/ark:/48223/pf0000387033_eng},
}

@online{weber2024emergency,
  author       = {Daniela Weber},
  title        = {AI in Emergency Response: Not the Time for Experiments},
  year         = {2024},
  organization = {NetHope - News and Press Releases},
  url          = {https://reliefweb.int/report/world/ai-emergency-response-not-time-experiments},
  note         = {Accessed: 2025-09-01}
}

@online{nethope2024empowering,
  author       = {NetHope - News and Press Releases},
  title        = {Empowering Humanitarian Response Through Crisis Informatics},
  year         = {2024},
  url          = {https://reliefweb.int/report/world/ai-emergency-response-not-time-experiments},
  note         = {Accessed: 2025-09-01}
}

@online{nethope2023amplifying,
  author       = {Daniela Weber},
  title        = {Amplifying the efforts of nonprofit organizations with AI},
  year         = {2023},
  organization = {NetHope - News and Press Releases},
  url          = {https://reliefweb.int/report/world/ai-emergency-response-not-time-experiments},
  note         = {Accessed: 2025-09-01}
}

@online{toplic2020ai,
  author       = {Leila Toplic},
  title        = {AI in the Humanitarian Sector},
  year         = {2020},
  organization = {NetHope - News and Press Releases},
  url          = {https://reliefweb.int/report/world/ai-emergency-response-not-time-experiments},
  note         = {Accessed: 2025-09-01}
}

@online {avagyan2020utilizing,
  author = {Avetis Avagyan and Hae-Yeon Alice Jeong},
  title = {Utilizing artificial intelligence for equitable and efficient volunteer selection},
  year = {2020},
  organization = {{UN} Volunteers},
  url = {https://www.unv.org/Success-stories/utilizing-artificial-intelligence-equitable-and-efficient-volunteer-selection},
  note = {Accessed: 2025-09-01}
}

@article{scutto2025empowerment,
title = {Tackling the empowerment of Artificial Intelligence for humanitarian intervention from Save the Children},
journal = {Technological Forecasting and Social Change},
volume = {221},
pages = {124330},
year = {2025},
issn = {0040-1625},
doi = {https://doi.org/10.1016/j.techfore.2025.124330},
author = {Veronica Scuotto and Kingsley Obi Omeihe and Valentina Cillo and Del Giudice Manlio},
}

@article{ahatsi2025resilience,
  title={Resilience in Humanitarian Supply Chains: Addressing Artificial Intelligence and Big Data Hurdles Across Borders},
  author={Ahatsi, Emmanuel and Olanrewaju, Oludolapo Akanni},
  journal={Engineering Reports},
  volume={7},
  number={7},
  pages={e70310},
  year={2025},
  publisher={Wiley Online Library},
  doi = {https://doi.org/10.1002/eng2.70310}
}

@article{iazzolino2025trading,
  title={Trading efficiency for control: The AI conundrum in migration management},
  author={Iazzolino, Gianluca},
  journal={Cosmopolitan Civil Societies: An Interdisciplinary Journal},
  volume={17},
  number={1},
  pages={35--46},
  year={2025},
  publisher={UTS ePress Sydney},
  doi = {https://doi.org/10.5130/ccs.v17.i1.9423}
}

@INPROCEEDINGS{bhuvaneswari2025unite,
author={Bhuvaneswari, R and K, Dhivya Shree and K, Divya Dharshini and H, Farheen Tabassum},
booktitle={2025 6th International Conference on Data Intelligence and Cognitive Informatics (ICDICI)}, 
title={UniteVol: AI-Powered Social Impact and Leadership Platform}, 
year={2025},
pages={1703-1707},
doi={10.1109/ICDICI66477.2025.11135302}
}

@article{popescu2024impact,
author = {Popescu, Cristina Raluca Gh and Duháček Šebestová, Jarmila},
year = {2024},
month = {09},
pages = {3899},
title = {The Impact of Artificial Intelligence on Intellectual Capital Development: Shifting Requirements for Professions and Processes in the Non-Profit Sector},
volume = {8},
journal = {Journal of Infrastructure Policy and Development},
doi = {10.24294/jipd.v8i10.3899}
}

@INPROCEEDINGS{uke2024food,
  author={Uke, Shailaja and Jadhav, Prithviraj and Kakarwal, Urmila and Kolawale, Sanika and Nikam, Vivek},
  booktitle={2024 International Conference on Sustainable Communication Networks and Application (ICSCNA)}, 
  title={FoodSavior: Distributing Surplus Food to NGOs or Manure Producers Using Internet of Things and Machine Learning}, 
  year={2024},
  pages={146-153}, 
  doi = {10.1109/ICSCNA63714.2024.10863911}
  }

@INPROCEEDINGS{gupta2024cyber,
  author={Gupta, Rohan and Singh, Ankit Kumar and Utkarsh and Mittal, Prachi and Radhika},
  booktitle={2024 3rd Edition of IEEE Delhi Section Flagship Conference (DELCON)}, 
  title={AI Based Cyberbullying Detection and Prevention}, 
  year={2024},
  pages={1-6},
  doi = {10.1109/DELCON64804.2024.10866680}
}

@INPROCEEDINGS{sanjana2024integrating,
author={K, Sanjana and R, Asha G and Vineeth, Nandhini},
booktitle={2024 2nd DMIHER International Conference on Artificial Intelligence in Healthcare, Education and Industry (IDICAIEI)}, 
title={Integrating Transparent Crowdfunding Platform and AI-Based Treatment Fund Estimation}, 
year={2024},
pages={1-6},
doi={10.1109/IDICAIEI61867.2024.10842939}}

@inproceedings{kbah2023understanding,
  title={Understanding Enablers and Barriers for Deploying AI/ML in Humanitarian Organizations: the Case of DRC's Foresight},
  author={Kbah, Zaid and Gralla, Erica},
  booktitle={Proceedings of the IISE Annual Conference \& Expo 2023},
  year={2023}
}

@article{rahman2021development,
title = {Development of flood hazard map and emergency relief operation system using hydrodynamic modeling and machine learning algorithm},
journal = {Journal of Cleaner Production},
volume = {311},
pages = {127594},
year = {2021},
issn = {0959-6526},
doi = {https://doi.org/10.1016/j.jclepro.2021.127594},
author = {Mahfuzur Rahman and Ningsheng Chen and Md Monirul Islam and Golam Iftekhar Mahmud and Hamid Reza Pourghasemi and Mehtab Alam and Md Abdur Rahim and Muhammad Aslam Baig and Arnob Bhattacharjee and Ashraf Dewan},
}


\appendix
\onecolumn

\clearpage
\section{Search Queries}
\label{appendix:search querys}
\begin{table*}[h!]
\centering
\caption{Initial search strings and results}
\label{tab:initial-search}
\begin{tabular}{lp{0.65\textwidth}lc}
\toprule
\textbf{Digital library} & \textbf{Search string} & \textbf{Search time} & \textbf{No. of} \\
& & & \textbf{papers}\\
\midrule

Google Scholar & 

("artificial intelligence" OR AI OR "machine learning" OR ML)
AND
("non-governmental organization*" OR NGO* OR "non-profit*" 
OR "charity*" OR "humanitarian*" OR "aid organization*") 
AND 
("social impact" OR "social good" OR "community development" 
OR "sustainability" OR "disaster response" OR "refugee support")

& 18/08/2025 & 18,300 \\
\addlinespace

OpenAlex &

'abstract.search:(("artificial intelligence" OR "machine learning" OR "AI") AND ("non-governmental organization" OR NGO OR "non-profit organization" OR "humanitarian organization")),'
'default.search:("social impact" OR "social good" OR "community development" OR "sustainability" OR "disaster response" OR "refugee support")'
),
& 18/08/2025 & 339 \\
\bottomrule
\end{tabular}
\end{table*}

\begin{table*}[h!]
\centering
\caption{Refined search strings and results}
\label{tab:refined-search}
\begin{tabular}{lp{0.65\textwidth}lc}
\toprule
\textbf{Digital library} & \textbf{Search string} & \textbf{Search time} & \textbf{No. of} \\
& & & \textbf{papers}\\
\midrule

Google Scholar &

(intitle:"artificial intelligence" OR intitle:"machine learning" 
OR intitle:AI OR intitle:"Natural language processing" OR intitle:NLP)   
AND    
(intitle:"non-governmental*" OR intitle:NGO OR intitle:"non-profit*" 
OR intitle:"humanitarian organization" OR intitle:"charity" 
OR intitle:"civil society organization") 
AND    
("humanitarian aid" OR "disaster response" OR "refugee support" 
OR "community development" OR "education" OR "public health" 
OR "development aid" OR "crisis management" OR "poverty alleviation" 
OR "sustainability" OR "human rights" OR "education")

& 18/08/2025 & 47 \\

Open Alex & 'title.search:("artificial intelligence" OR "machine learning" OR "AI"),'
'abstract.search:("non-governmental organization" OR NGO 
OR "non-profit organization" OR "humanitarian organization"),'
'default.search:("humanitarian aid" OR "disaster response" 
OR "refugee support" OR "community development" OR "education" 
OR "public health" OR "development aid" OR "crisis management" 
OR "poverty alleviation" OR "sustainability" OR "human rights")'

& 18/08/2025 & 129 \\

Scopus & Title: ("artificial intelligence" OR "machine learning" OR "AI") AND Abstract: ("non-governmental organization" OR NGO OR "non-profit organization" OR "humanitarian organization") AND All fields: ("humanitarian aid" OR "disaster response" OR "refugee support" OR "community development" OR "education" OR "public health" OR "development aid" OR "crisis management" OR "poverty alleviation" OR "sustainability" OR "human rights") 

& 01/10/2025 & 91\\

Relief Web & ("artificial intelligence" OR "machine learning" OR "AI")  AND ("adoption" OR "implementation" OR "strategy")  AND ("NGO" OR "non-governmental organization" OR "nonprofit" OR "NPO" OR "charity" OR "aid organization") \textit{[using filter for Non-Governmental Organization]}& 18/08/2025 & 83 \\
\bottomrule
\end{tabular}
\end{table*}

\clearpage
\section{Thematic Map}
\label{app:thematic map}
\begin{figure*}[h!]
    \centering
    \includegraphics[width=0.9\textwidth]{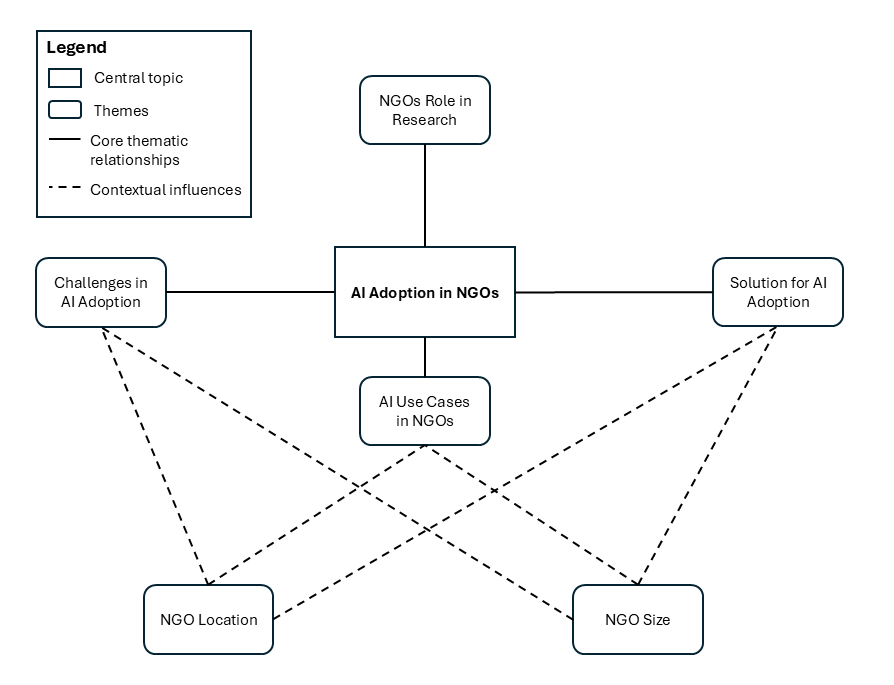} 
    \caption{Thematic mapping. Note that each thematic category contains multiple sub-themes. For reasons of visual clarity, only the main themes are displayed here. For a mapping of used sub-themes to concrete papers please refer to the \href{https://osf.io/a2357/?view_only=7120b178719540608dc97cd2a178ceff}{additional material}.}
    \label{fig:pipeline}
\end{figure*}

\clearpage
\section{Overview of Included Studies}
\label{app:overview}

\begin{table*}[h!]
  \caption{Included studys from 2020}
  \label{tab:overview}
  \begin{tabular}{p{3cm} p{6cm} l p{3cm} c}
    \toprule
    Authors & Title & Type & Discussed Topics & Citation\\
    \midrule
    Tingzon et al. & Mapping New Informal Settlements Using Machine Learning and Time Series Satellite Images: An Application in the Venezuelan Migration Crisis & Conference paper & Use Cases, Solutions & \cite{tingzon2020mapping}\\
    Defnizal \& Ernes & The Implementation of Artificial Intelligence in Charity Box at Mosque and Musholla as RFID Based Security System & Article & Use cases & \cite{defnizal2020ai}\\
    Alhindi et al. & Vehicle routing optimization for surplus food in nonprofit organizations & Article & Use cases, Solutions & \cite{alhindi2020vehicle}\\
    Toplic & AI in the Humanitarian Sector & NGO paper & Use Cases, Challenges, Solutions & \cite{toplic2020ai}\\
    Avagyan \& Jeong & Utilizing Artificial Intelligence for Equitable and Efficient Volunteer Selection & NGO paper & Use cases, Solutions & \cite{avagyan2020utilizing}\\
    \bottomrule
  \end{tabular}
\end{table*}

\begin{table*}[h!]
  \caption{Included studys from 2021}
  \label{tab:overview}
  \begin{tabular}{p{3cm} p{6cm} l p{3cm} c}
    \toprule
    Authors & Title & Type & Discussed Topics & Citation\\
    \midrule
    Ghani & Machine Learning for Social Good: Applications in Non-Profit and Public Sectors & Article & Use cases, Challenges, Solutions & \cite{ghani2021ml}\\
    Rahman et al. & Development of flood hazard map and emergency relief operation system using hydrodynamic modeling and machine learning algorithm & Article & Use cases, Solutions & \cite{rahman2021development}\\
    Sharma et al. & OCR for Data Retrieval: An analysis and Machine Learning Application model for NGO social volunteering & Conference paper & Use cases, Challenges & \cite{sharma2021ocr}\\
    \bottomrule
  \end{tabular}
\end{table*}

\begin{table*}[h!]
  \caption{Included studys from 2022}
  \label{tab:overview}
  \begin{tabular}{p{3cm} p{6cm} l p{3cm} c}
    \toprule
    Authors & Title & Type & Discussed Topics & Citation\\
    \midrule
    Dubey et al. & Impact of artificial intelligence-driven big data analytics culture on agility and resilience in humanitarian supply chain: A practice-based view & Article & Use Cases, Challenges & \cite{dubey2022impact}\\
    Castro et al. & AI + Dance: Co-Designing Culturally Sustaining Curricular Resources for AI and Ethics Education Through Artistic Computing & Conference paper & Use Cases & \cite{castro2022ai}\\
    Rathore et al. & A Sustainable Model for Emergency Medical Services in Developing Countries: A Novel Approach Using Partial Outsourcing and Machine Learning & Article & Use Cases & \cite{rathore2022sustainable}\\
    Tabar et al. & Forecasting the Number of Tenants At-Risk of Formal Eviction: A Machine Learning Approach to Inform Public Policy & Article & Use cases, Solutions & \cite{tabar2022forecasting}\\
    Nair et al. & ADVISER: AI-Driven Vaccination Intervention Optimiser for Increasing Vaccine Uptake in Nigeria & Conference Paper & Use cases, Challenges, Solutions & \cite{nair2022adviser}\\
    Alnamrouti et al. & Do Strategic Human Resources and Artificial Intelligence Help to Make Organisations More Sustainable? Evidence from Non-Governmental Organisations & Article & Use cases, Challenges, Solutions & \cite{alnamrouti2022ai}\\
    Mate et al. & Field study in deploying restless multi-armed bandits: Assisting non-profits in improving maternal and child health & Conference paper & Use cases, Challenges, Solutions & \cite{mate2022bandits}\\
    \bottomrule
  \end{tabular}
\end{table*}

\begin{table*}[h!]
  \caption{Included studys from 2023}
  \label{tab:overview}
  \begin{tabular}{p{3cm} p{6cm} l p{3cm} c}
    \toprule
    Authors & Title & Type & Discussed Topics & Citation\\
    \midrule
    Efthymiou et al. & The Role of Artificial Intelligence in Revolutionizing NGOs' Work & Article & Use Cases, Challenges, Solutions & \cite{efthymiou2023role}\\
    Grass et al. & A machine learning approach to deal with ambiguity in the humanitarian decision making & Article & Use Cases, Challenges, Solutions & \cite{grass2023ml}\\
    Bahameish et al. & Artificial Intelligence in Procurement: An Overview and Case Study of Qatar Foundation & Conference paper & Use Cases, Challenges & \cite{bahameish2023ai}\\
    Soudi et al. & Generative AI-Based Tutoring System for Upper Egypt Community Schools & Conference paper & Use cases, Challenges, Solutions & \cite{soudi2023generative}\\
    Niranjana & AI for Sustainable Development: Assessing Student Interest, Education, and Career Pathways & Article & Challenges & \cite{niranjana2023ai}\\
    Alex \& Sotiris & Designing Artificial Intelligence Equipped Social Decentralized Autonomous Organizations for Tackling Sextortion Cases Version 0.7 & Preprint & Use Cases, Challenges, Solutions & \cite{alex2023designing}\\
    Rafner et al. & Crea.visions : A Platform for Casual Co-Creation with a Purpose Envisioning the Future through Human-AI Collaboration with Multiple Stakeholders & Conference Paper & Use cases & \cite{rafner2023crea}\\
    Laylo & The Impact of AI and Information Technologies on Islamic Charity (Zakat): Modern Solutions for Efficient Distribution & Article & Use cases & \cite{laylo2023impact}\\
    Pai et al. & NGO CONNECT: Technology for Non-Profit Organisation Management & Conference paper & Use cases, Challenges, Solutions & \cite{pai2023ngo}\\
    Arango et al. & Consumer Responses to AI-Generated Charitable Giving Ads & Article & Challenges, Solutions & \cite{arango2023aiads}\\
    Khan \& Savolainen & Sal Khan: "I see AI as an additional tool, but a very powerful one" & Article & Challenges, Solutions & \cite{unesco2023sal}\\
    Kbah \& Gralla & Understanding Enablers and Barriers for Deploying AI/ML in Humanitarian Organizations: the case of DRC’s Foresight & Conference paper & Use cases, Challenges, Solutions & \cite{kbah2023understanding}\\
    Nethope (org.) & Amplifying the efforts of nonprofit organizations with AI & NGO paper & Challenges, Solutions & \cite{nethope2023amplifying}\\
    \bottomrule
  \end{tabular}
\end{table*}

\begin{table*}[h!]
  \caption{Included studys from 2024 (Part 1)}
  \label{tab:overview}
  \begin{tabular}{p{3cm} p{6cm} l p{3cm} c}
    \toprule
    Authors & Title & Type & Discussed Topics & Citation\\
    \midrule
    Pereira \& Shafique & The Role of Artificial Intelligence in Supply Chain Agility: A Perspective of Humanitarian Supply Chain & Article & Use Cases, Solutions & \cite{pereira2024ai}\\
    Huang & Technology-Driven Financial Risk Management: Exploring the Benefits of Machine Learning for Non-Profit Organizations & Article & Use Cases, Challenges & \cite{huang2024tech}\\
    Casagran \& Stavropoulos & Developing AI predictive migration tools to enhance humanitarian support: The case of EUMigraTool & Article & Use cases, Challenges, Solutions & \cite{blasi2024eumigratool}\\
    Asajile et al. & Influence of Artificial Intelligence on Selection Stage of Recruitment in Tanzania: A Case of Selected NGOs in Kinondoni Municipality & Article & Use Cases, Challenges, Solutions & \cite{asajile2024ai}\\
    Thalor et al. & Green Mapper: An AI-Driven Initiative for Aerial Tree Mapping, Maintaining Environmental Balance & Article & Use cases, Solutions & \cite{thalor2024greenmapper}\\
    Dube et al. & Factors Influencing the Adoption of AI Chatbots By Non-Governmental Organizations & Conference Paper & Use cases, Challenges, Solutions & \cite{dube2024factors}\\
    Faruq et al. & AI-Driven Strategies for Enhancing Non-Profit Organizational Impact & Article & Use cases, Challenges, Solutions & \cite{faruq2024ai}\\
    Kapuge \& Ginige & Optimizing AI Recommendation Algorithms for Efficient Matching of Most Needed Beneficiaries with Donors in Sri Lankan Charity Sector & Conference paper & Use cases & \cite{kapuge2024optimizing}\\
    Sammer et al. & AI-FEED: Prototyping an AI-Powered Platform for the Food Charity Ecosystem & Article & Use cases, Solutions & \cite{sammer2024aifeed}\\
    Sievert \& Inderhees & AI, Diversity and Trust in Digital Non-Profit Communication: Results of a Survey within the Major Traditional Churches of a Western European Country & Conference paper & Use cases & \cite{sievert2024ai}\\
    Dutta \& Kannan Poyil & The machine/human argentic impact on practices in learning and development: a study across MSME, NGO and MNC organizations & Article & Use cases, Challenges & \cite{dutta2024machine}\\
    Elamin & Modernizing the Charitable Sector through Artificial Intelligence: Enhancing Efficiency and Impact & Article & Use cases, Challenges, Solutions & \cite{elamin2024ai}\\
    Yang et al. & Understanding the impact of artificial intelligence on the justice of charitable giving: The moderating role of trust and regulatory orientation & Article & Challenges & \cite{yang2024aijustice}\\
    Popescu et al. & The impact of artificial intelligence on intellectual capital development: Shifting requirements for professions and processes in the non-profit sector & Article & Use cases, Challenges, Solutions & \cite{popescu2024impact}\\
    \bottomrule
  \end{tabular}
\end{table*}

\begin{table*}[h!]
  \caption{Included studys from 2024 (Part 2)}
  \label{tab:overview}
  \begin{tabular}{p{3cm} p{6cm} l p{3cm} c}
    \toprule
    Authors & Title & Type & Discussed Topics & Citation\\
    \midrule
    Gupta et al. & AI Based Cyberbullying Detection and Prevention & Conference paper & Use cases, Challenges & \cite{gupta2024cyber}\\
    Sanjana et al. & Integrating Transparent Crowdfunding Platform and AI-Based Treatment Fund Estimation & Conference paper & Use cases & \cite{sanjana2024integrating}\\
    Weber & AI in Emergency Response: Not the time for experiments & NGO paper & Use cases, Challenges, Solutions & \cite{weber2024emergency}\\
    Nethope (org.) & Empowering Humanitarian Response Through Crisis Informatics & NGO paper & Challenges, Solutions & \cite{nethope2024empowering}\\
    Uke et al. & FoodSavior: Distributing Surplus Food to NGOs or Manure Producers Using Internet of Things and Machine Learning & Conference paper & Use cases & \cite{uke2024food}\\
    \bottomrule
  \end{tabular}
\end{table*}

\begin{table*}[h!]
  \caption{Included studys from 2025 (Part 1)}
  \label{tab:overview}
  \begin{tabular}{p{3cm} p{6cm} l p{3cm} c}
    \toprule
    Authors & Title & Type & Discussed Topics & Citation\\
    \midrule
    Brezovar & The Role of Artificial Intelligence in NGOs: Challenges and Opportunities for Slovenia's Information Society & Article & Use Cases, Challenges, Solutions & \cite{nejc2025role}\\
    Victoire & A Machine Learning Approach to Categorizing Countries by Socio-economic and Health Development Factors Using PCA, K-Means, and Silhouette Scoring & Article & Use Cases, Challenges & \cite{amalraj2025ml}\\
    Atalay et al. & Artificial Intelligence Technologies as Smart Solutions for Sustainable Protected Areas Management & Article & Use cases, Challenges, Solutions & \cite{atalay2025ai}\\
    Ahatsi \& Olanrewaju & Enhancing Humanitarian Supply Chain Resilience: Evaluating Artificial Intelligence and Big Data Analytics in Two Nations & Article & Use Cases, Challenges, Solutions & \cite{ahatsi2025humanitarian}\\
    Ulfy et al. & Integration of Artificial Intelligence in Biodegradable Plastic Packaging Design: Exploring Stakeholder Attitudes & Article & Use cases, Challenges, Solutions & \cite{ulfy2025ai}\\
    Kravchuk & Ethical Implications of AI Applications in Nonprofit and Charity Sectors & Article & Use Cases, Solutions, Challenges & \cite{kravchuk2025ethical}\\
    Kazemeini et al. & A Dataset-Driven Study of AI Opportunities in the Climate NGO Ecosystem & Conference paper & Use cases, Challenges & \cite{kazemeini2025dataset}\\
    Krause & AI Agents and Automation in Small Non-Profit Organizations' Accounting Functions & Article & Use cases, Challenges, Solutions & \cite{krause2025ai}\\
    Kannan et al. & AI-Driven Digital Volunteering @ Freelancing Website for NGO Collaboration with Online Talent Worldwide & Article & Use cases & \cite{kannan2025ai}\\
    Hansen & Exclusion or Efficiency: Understanding Perspectives about AI Ethics Among Charity Workers in the United Kingdom & Conference paper & Use cases, Challenges & \cite{hansen2025ai}\\
    \bottomrule
  \end{tabular}
\end{table*}

\begin{table*}[h!]
  \caption{Included studys from 2025 (Part 2)}
  \label{tab:overview}
  \begin{tabular}{p{3cm} p{6cm} l p{3cm} c}
    \toprule
    Authors & Title & Type & Discussed Topics & Citation\\
    \midrule
    Cheng \& Wang & Leveraging Artificial Intelligence–Powered Chatbots for Nonprofit Organizations: Examining the Antecedents and Outcomes of Chatbot Trust and Social Media Engagement & Article & Use cases, Challenges, Solutions & \cite{cheng2025aifeed}\\
    Hahn et al. & From Data to Donors: Can AI Reshape Fundraising Strategies & Conference paper & Use cases, Challenges & \cite{hahn2025strategic}\\
    Joshi et al. & SmartNGO: An Integrated Platform for Managing Volunteers and Events & Article & Use cases & \cite{joshi2025smart}\\
    Scuotto et al. & Tackling the empowerment of Artificial Intelligence for humanitarian intervention from Save the Children & Article &  Use cases, Challenges & \cite{scutto2025empowerment}\\
    Ahatsi et al. & Resilience in Humanitarian Supply Chains: Addressing Artificial Intelligence and Big Data Hurdles Across Borders & Article & Use cases, Challenges, Solutions & \cite{ahatsi2025resilience}\\
    Iazzolino & Trading Efficiency for Control: the AI Conundrum in Migration Management & Article &  Use cases, Challenges, Solutions & \cite{iazzolino2025trading}\\
    Bhuvaneswari et al. & UniteVol: AI-Powered Social Impact and Leadership Platform & Conference paper & Use cases & \cite{bhuvaneswari2025unite}\\
    Sandberg et al. & Addressing the promise and peril of AI for nonprofit management through a data feminist pedagogy & Article & Use cases, Challenges, Solutions & \cite{sandberg2025ai}\\
    \bottomrule
  \end{tabular}
\end{table*}

\end{document}